\def\kms{\ifmmode{\,\hbox{km}\,s^{-1}}\else {\rm\,km\,s$^{-1}$}\fi}

\def\msun{{\rm\,M_\odot}}
\def\lsun{{\rm\,L_\odot}}

\def\kmsm{{\rm\,km\,s^{-1}\,Mpc^{-1}}}

\def\hmpc{\ifmmode{h^{-1}\,\hbox{Mpc}}\else{$h^{-1}$\thinspace Mpc}\fi}
\def\hkpc{\ifmmode{\,h^{-1}\,{\rm kpc}}\else {$h^{-1}$\,kpc}\fi}

\def\et{{\it et~al.}~}

\documentclass[12pt,preprint]{aastex}
\begin{document}
\tolerance 8000
%\slugcomment{DRAFT: \today}

\title{Environment and Galaxy Evolution at Intermediate Redshift 
in the CNOC2 Survey}

\author{R.~G.~Carlberg\altaffilmark{1,2},
H.~K.~C.~Yee\altaffilmark{1,2}, S.~L.~Morris\altaffilmark{1,3,4},
H.~Lin\altaffilmark{1,2,5,6}, \\ P.~B.~Hall\altaffilmark{1,2,7},
D.~R.~Patton\altaffilmark{1,2,8}, M.~Sawicki\altaffilmark{1,2,9}, and
C.~W.~Shepherd\altaffilmark{1,2} }
\altaffiltext{1}{Visiting Astronomer, Canada--France--Hawaii Telescope, 
        which is operated by the National Research Council of Canada,
        le Centre National de Recherche Scientifique, and the University 
        of Hawaii.}
\altaffiltext{2}{Department of Astronomy, University of Toronto, 
        Toronto ON, M5S~3H8 Canada}
\altaffiltext{3}{Dominion Astrophysical Observatory, 
        Herzberg Institute of Astrophysics,    ,  
        National Research Council of Canada,
        5071 West Saanich Road,
        Victoria, BC, V8X~4M6, Canada}
\altaffiltext{4}{Department of Physics, University of Durham,
	South Road, Durham DH1 3LE, UK}
\altaffiltext{5}{Steward Observatory, University of Arizona,
        Tucson, AZ, 85721}
\altaffiltext{6}{Hubble Fellow}
\altaffiltext{7}{Princeton University Observatory, Princeton, NJ 08544-1001 and
Pontificia Universidad Cat\'{o}lica de Chile, Departamento de
Astronom\'{\i}a y Astrof\'{\i}sica, Facultad de F\'{\i}sica, Casilla
306, Santiago 22, Chile}
\altaffiltext{8}{Department of Physics \& Astronomy,
        University of Victoria,
        Victoria, BC, V8W~3P6, Canada}
\altaffiltext{9}{Mail Code 320-47, Caltech, Pasadena 91125, USA}

%\clearpage

\begin{abstract}
The systematic variation of galaxy colors and types with clustering
environment could either be the result of local conditions at
formation or subsequent environmental effects as larger scale
structures draw together galaxies whose stellar mass is largely in
place.  Below redshift 0.7 galaxy luminosities (k-corrected and
evolution compensated) are relatively invariant, whereas galaxy star
formation rates, as reflected in their colors, are a ``transient''
property that have a wide range for a given luminosity. The relations
between these galaxy properties and the clustering properties are key
statistics for understanding the forces driving late-time galaxy
evolution.  At $z\sim0.4$ the co-moving galaxy correlation length,
$r_0$, measured in the CNOC2 sample is strongly color dependent,
rising from 2\hmpc\ to nearly 10\hmpc\ as the volume-limited
subsamples range from blue to red.  The luminosity dependence of $r_0$
at $z\sim0.4$ is weak below $L_\ast$ in the R band, although there is
an upturn at high luminosity where its interpretation depends on
separating it from the $r_0$-color relation. In the B band there is a
slow, smooth increase of $r_0$ with luminosity, at least partially
related to the color dependence.  Study of the evolution of galaxies
within groups, which create much of the strongly nonlinear correlation
signal, allows a physical investigation of the source of these
relations.  The dominant effect of the group environment on star
formation is seen in the radial gradient of the mean galaxy colors
which on the average become redder than the field toward the group
centers. The color differentiation begins around the dynamical radius
of virialization of the groups.  The redder-than-field trend applies
to groups with a line-of-sight velocity dispersion, $\sigma_1>150
\kms$. There is an indication, somewhat statistically insecure, that
the high luminosity galaxies in groups with $\sigma_1<125 \kms$ become
bluer toward the group center.  Monte Carlo orbit integrations
initiated at the measured positions and velocities show that the rate
of galaxy merging in the $\sigma_1>150\kms$ groups is very low,
whereas for $\sigma_1<150\kms$ about 25\% of the galaxies will merge
in 0.5Gyr. We conclude that the higher velocity dispersion groups
largely act to suppress star formation relative to the less clustered
field, leading to ``embalmed'' galaxies. On the other hand, the low
velocity dispersion groups are prime sites of both strong merging and
enhanced star formation that leads to the formation of some new
massive galaxies at intermediate redshifts.  The tidal fields within
the groups appear to be a strong candidate for the physical source of
the reduction of star formation in group galaxies relative to
field. Tides operate effectively at all velocity dispersions to remove
gas rich companions and low density gas in galactic halos.  We find a
close resemblance of the color dependent galaxy luminosity function
evolution in the field and groups, suggesting that the clustering
dependent star formation reduction mechanism is important for the
evolution of field galaxies as a whole.
\end{abstract}

\keywords{cosmology: large scale structure, galaxies: evolution}

\section{Introduction}

Environment plays a major role in the formation and evolution of
galaxies as is dramatically manifest in the differences between
galaxies in rich clusters and the field \citep{dressler}.  The
environmental effect is mainly one to alter the distribution over
galaxy type, with even the somewhat exotic E+A galaxies being found in
the field \citep{zab_ea}. A notable exception is the cD galaxies,
but even those are believed to originate from an early-epoch field population
\citep{dubinski_cd}. It is ambiguous at any single redshift 
whether the systematically redder and earlier morphological types of
cluster galaxies, as compared to the field galaxy population, are the
result of different local conditions at the time of formation, or,
whether all galaxies begin as a fairly uniform distribution of types
which subsequent evolutionary forces differentiate as a function of
environment.  Examining the redshift dependence of galaxy populations
is a powerful tool to separate low redshift environmentally driven
changes from high redshift formation mechanisms as the primary cause
of the differences.

An example of the study of the environmental dependence of populations
with redshift is the star formation rate of cluster galaxies at
intermediate redshift. On the average, cluster galaxy star formation
rates are smaller than those in field galaxies, with a progressive
decline to the cluster center
\citep{dressler,whitmore,balogh,poggianti,pca}.  This suggests that a
field ``mix'' of galaxies is capable of being transformed into an
earlier type cluster population in response to the cluster
environment. Cluster galaxies provide important clues of environmental
forcing but they are a moderately rare population subject to
exceptionally strong environmental forces.

Galaxy groups show many of the differences from the field galaxy
population that cluster galaxies exhibit
\citep{zab_gps,ho,tran_gps}.  The evolution of group galaxies is
necessarily a significant component of the evolution of the entire
galaxy population.  Approximately $20\%$ of $L\ga L_\ast/3$ galaxies
reside in small virialized galaxy groups
\citep{nw,cnoc2_gps}, and virtually all galaxies are in groups if the
luminosity limit for inclusion is lowered sufficiently.  For instance,
the Milky Way Galaxy, with the Large and Small Magellanic Clouds,
constitutes a small virialized group (which falls just slightly short
of meeting the luminosity requirements of the groups we define
below). The Local Group also usefully illustrates that the 
lowest
velocity dispersion groups overlap the velocity dispersion range of
individual and pairs of galaxies.  The mean pairwise 1D peculiar
velocity dispersion of galaxies is 300-500 \kms\
\citep{dp,marzke,2df_vel,sdss_clus} indicating that groups with one
dimensional velocity dispersions, $\sigma_1$, in the range of 200-300
\kms\ dominate the clustering of field galaxies.  

Some combination of internal clocks and external agents drives galaxy
evolution. Internal timescales are regulated by dynamical and thermal
processes, with generally fairly fast clocks, whereas the external
timescales are controlled by the structures in which galaxies find
themselves and the rate of development of large scale clustering,
clocks running at approximately the Hubble rate.  At a given galaxy
type, both field and group galaxies have essentially identical
internal processes. Consequently, clustering dependent population
differences must be due to either conditions at formation or
subsequent environmental interactions. Galaxy formation is not a
process that occurs at a sharply defined time.  However, for the
intermediate redshift range we examine here normal luminous galaxies
of a wide range of type can be taken to essentially ``formed'' with
varying degrees of ongoing star formation and merging that do not
grossly alter the stellar mass of the galaxies.

Gravitational tides, dynamical friction and intergalactic gas pressure
are two physical forces present in groups that can affect their galaxy
contents. Small galaxy groups have been investigated extensively at
low redshift
\citep{bb,cfa_gps,cfa_newgps,mahdavi} and at intermediate redshift 
\citep{cnoc2_gps} to show that the total mass of a virialized group is
dominated by dark matter. One action of the dark matter is to cause
dynamical friction, whose in-spiral timescale varies in proportion to
$\sigma^3/\rho$.  Detailed dynamical analysis
\citep{merritt_ldist,merritt_cd} and n-body simulations
\citep{mamon,dubinski_cd} provide quantitative evidence
that high velocity dispersion clusters do not suffer much internal
dynamical evolution of their galaxy populations after their primary
formation phase. However, low velocity dispersion groups are found to
produce mergers on a cosmologically short timescale, even at low
redshift where the virialization density, approximately 200 times the
critical density, is at a minimum
\citep{barnes,barnes_emg,bcl,bbcl,mamon,diaferio,harassment}. Galaxy groups
also contain X-ray emitting hot gas that work on the gas content of
galaxies \citep{mulchaey_review}.  Ram pressure stripping declines in
proportion to the velocity dispersion (squared), so plays a relatively
less important role in groups than in large clusters
\citep{abadi,balogh_model}.

The evidence that groups play a significant role in galaxy evolution
is strong, but the observationally confirmed effects of groups vary with
the group definition and are largely confined to low redshift
samples. For instance, the Hickson (1982) compact groups are
deliberately designed to be quite high density groups.  They are found
to include a number of strongly interacting and likely-to-be-merging
systems with enhanced star formation \citep{hickson_rev,ho}. The
ultimate fate predicted for the galaxies in isolated low velocity
dispersion groups is to join to form a single massive elliptical
galaxy \citep{barnes_emg,mz_e}. Groups defined with a lower density
threshold are observed to have suppressed star formation relative to
the field
\citep{lcrs_gps,tran_gps} reminiscent of the effect that is seen in rich
clusters.

An empirical study of groups uniformly selected from the field at
$z\sim 0.5$ opens up the possibility of detecting differential
evolutionary effects between groups and the field.  Galaxy evolution
over the range from $z=0$ to $z\sim 0.7$ is substantial
\citep{bes} particularly in  relatively blue galaxies less luminous
than $L_\ast$ \citep{cfrs_lf,huan_lf}. The physical origin of this
evolution remains uncertain, although it is clear that in indvidual
galaxies it is superimposed on a slowly aging underlying stellar
population and that it primarily requires regulation of the star
formation rate
\citep{lilly_sfr}. Galaxy groups in this redshift range have been
defined with photometric data \citep{demello} and with radio-galaxies
as centers \citep{radio_gps} but large, kinematically accurate
redshift catalogs are needed to put intermediate redshift groups on
the same standing as those at low redshift.

In the next section we review the Canadian Network for Observational
Cosmology field galaxy redshift survey (CNOC2) sample which contains
about 6000 galaxies with velocities measured to a precision of about
70-100 \kms\ over the $0.1-0.55$ redshift range. The sample was
selected to minimize clustering dependent differential selection
effects so that the relative evolution of group and field galaxies can
be readily examined.  We first use the field sample as a whole to show
the correlation function's strong dependence on color and relatively
weak luminosity dependence.  Section 3 discusses the color and
luminosity distribution of field and group galaxies and their
change with redshift. The variation of the mean galaxy colors with 
distance from the group center is discussed for high and low velocity
dispersion groups in Section 4. The orbits of the group galaxies are
Monte Carlo integrated in Section 5 to determine their likelihood of
merging. Section 6 discuses the implications of our findings for
galaxy evolution and states our conclusions. Most quantities in this
paper are derived with $H_0=100h
\kmsm$ and an $\Omega_M=0.2$, $\Omega_\Lambda=0$ cosmology, with
exceptions noted.

\section{The Galaxy Clustering-Color Relation at $z\sim 0.4$}

The CNOC2 field galaxy redshift survey covers approximately 1.5 square
degrees of sky spread over 4 widely separated sky patches
\citep{cnoc2_tech}. The unbiased redshift
sample covers the 0.1 to 0.55 redshift range. The survey has an
accurately measured selection function that varies from nearly 100\%
at R=19 mag down to about 20\% at the limit of R=21.5 mag. The
selection function is used to correct the spectroscopic sample numbers
back to what a complete sample would, on the average, contain. For each
galaxy k-corrections are evaluated in all observed colors using the
closest approximating Coleman, Wu \& Weedman (1980) spectral energy
distribution (SED) using a finely interpolated set of SEDs. The
evolution of the k-corrected galaxy luminosity function as a function
of non-evolving SED type has been discussed elsewhere
\citep{huan_lf}. We use these results to define a luminosity evolution
compensation term, intended to identify statistically the same
population over our entire redshift range. This is valid for the
population as a whole or two color sub-classes which are our samples of
interest.  An accurate galaxy-by-galaxy correction requires more
information, ideally using a mass indicator such as IR luminosity or
dynamical masses.  The k-corrected and evolution compensated R band
luminosities, $M_R^{ke}$, are calculated as $M_R-k_R(z)+E_z z$, where
the mean evolution rate, $E_z$, is approximately compensated with a
fading of one magnitude per unit redshift for all SED types. This
approach is statistically acceptable for the relatively high
luminosity galaxies we use in this paper, but cannot be extended to
lower luminosity blue galaxies which require either an allowance for
more luminosity evolution or number evolution.

The central observational results presented in this paper address the
dependence of colors on clustering environment. Within our photometric
dataset the color which best serves as our star formation rate
indicator is the k-corrected color $(B-R)_0 = (B-R)-k_B+k_R$.  We are
motivated to use this color because all our galaxies have $B$ and $R$
observations, with less deep coverage in $U$ and a limited redshift
range for the [OII] 3727\AA\ line. Of course $(B-R)_0$ has a
dependence on factors other than star formation alone, such as
metallicity, dust and the details of the star formation history. Our
study depends on a comparison of various sub-samples where $(B-R)_0$
serves to indicate the relative star formation rates.

To explicitly relate the colors to star formation rates we use the
PEGASE.2 models \citep{pegase}.  We also examine the stellar mass-to-light
ratio, $M/L$, for its population dependence. Figure~\ref{fig:SFR_BR}
plots the star formation rate and the mass-to-light ratio, M/L, in the
R and B bands, against the rest-frame $B-R$ color. Points are plotted
every $3\times 10^8$ model years of age.  All the photometric models
are assumed to be dust free and have exponentially declining star
formation rates, $e^{-t/\tau}$, with $\tau=1$ 2, 4 and 6 Gyr at
solar metallicity and a $\tau=1$ Gyr at twice solar. The colors
are plotted for model ages of 3 to 12 Gyr.  The figure shows that
$(B-R)_0$ color (assuming no dust) indicates the current star formation
rate, largely independent of the details of the history of the star
formation, although we see the well known effect that the metallicity
is important in the color of galaxies with low star formation
rates. We note that the M/L value of the stellar population varies
about a factor of two less in the R band than in the B band.

Our analysis of the redshift dependence of the two-point correlation
function of the CNOC2 sample finds that the correlation length,
$r_0(z)$, is nearly constant in co-moving co-ordinates
\citep{cnoc2_xi}. The sample is volume limited on the basis of
k-corrected and evolution compensated luminosities, $M_R^{ke}\le -20$
mag. Our correlation measurements are consistent with the theoretical
predictions of a slow increase of biasing of galaxies relative to the
dark matter increasing with redshift.

The intermediate redshift clustering of the CNOC2 sample varies
strongly with the mean SED type of the galaxies \citep{chuck_xi}, with
late type galaxies being far less clustered than early type
galaxies. The clustering analysis divides galaxies based
on non-evolving SED types, hence, the dependence of clustering
with type evolves with redshift as galaxies move to later types with
increasing redshift \citep{chuck_xi}.  Here we complement our earlier
analysis with a larger sample, in a narrower redshift range. We are
particularly interested in the relative sensitivity of $r_0$ to star
formation and luminosity. We use the redshift range 0.25 to 0.45, in
which the bulk of the CNOC2 sample falls. Figure~\ref{fig:xiBR} shows
the dependence of the co-moving correlation length on $(B-R)_0$,
analyzed in a flat $\Omega_M=0.2, \Omega_\Lambda=0.8$ cosmology
($\Lambda=0$ gives $r_0$ that are about 10\% smaller at the average
redshift of 0.39).  The sample is volume limited, with $M_R^{k,e}\le
-18.5$ mag. The correlation measurements are made with exactly the
same procedures as discussed in Carlberg \et\ (2000). We fit the
measured projected correlation function to the projection of a power
law correlation function, $(r_0/r)^\gamma$, where all lengths are
measured in co-moving units.  The derived $r_0$ and $\gamma$ are shown
in Figure~\ref{fig:xiBR} along with the errors estimated on the basis
of the variance of the four observational patches in the sample.  The
slope of the correlation function, $\gamma$ becomes steeper with
increasingly red color.  Both of these effects are known at lower
redshift as morphological type or emission/absorption line galaxy
correlations although the large range of correlation amplitudes we
find is not so readily apparent at low redshift
\citep{dg_morph,saunders, fisher,guzzo,willmer,loveday_apm}.  The most
relevant measure of correlation amplitude in a situation where the
$\gamma$ vary systematically is to show the correlation length
normalized to a fixed $\gamma$, here chosen to be $\gamma=1.8$.  As
shown in Figure~\ref{fig:xiBR} these values range from about 2\hmpc\
for the bluest galaxies to nearly 10\hmpc\ for the reddest galaxies. The
errors in these figures are estimated using the patch to patch variance over
our four widely separated patches, which is a robust empirical estimator.

Shepherd et al. (2001) found that there is a weak luminosity increase
(or none) of clustering within each SED class over a luminosity range
of 1.5 magnitudes.  Here we examine the luminosity dependence of
clustering over a 3.0 magnitude range, displaying the result in
Figure~\ref{fig:xilum}. Recall that in the R band
$M^{ke}_\ast(z=0.3)\simeq -20.3$ mag and in the B band
$M^{ke}_\ast(z=0.3)\simeq -19.1$ mag.  At lower luminosities,
$0.2L_\ast$ to $2L_\ast$, the Figure shows that $r_0$ could rise about
10\% for the factor of ten luminosity increase (for $\gamma=1.8$), but
the difference is statistically consistent with an even weaker
relation. Galaxies more luminous than $2L_\ast$ show a highly
significant increase in clustering, rising from $4\hmpc$ for $L\sim
L_\ast$ to 8\hmpc.  However, the mean color in this bin is
$(B-R)_0=1.55$ mag, which is about 0.25 mag redder than the lower
luminosity members of the sample.  Referring to the color dependence
of Figure~\ref{fig:xiBR} indicates that at a mean color of
$(B-R)_0=1.7$ we expect $r_0\simeq 7.2\pm 0.6\hmpc$. The highest
luminosity galaxies here have $r_0\simeq 8.0\pm
0.7\hmpc$. Consequently the increase in clustering in this high
luminosity sample could be largely the result of the correlation-color
relation. This is difficult to disentangle, since virtually all high
luminosity galaxies are quite red, leaving too few luminous blue
galaxies in the present sample for an accurate clustering measurement.

We can compare our intermediate redshift results with similar low 
redshift results.  We show in Figure~\ref{fig:xilum}a the 
clustering-luminosity relation found in the low redshift SDSS 
\citep{sdss_clus} using R band luminosity, where we convert SDSS $r_\ast$
to Cousins $R$ using $R=r_\ast-0.24$
\citep{SDSS_colors}.  A similar low redshift R band luminosity dependence 
of clustering was seen in the LCRS \citep{lcrs_power}.  In
Figure~\ref{fig:xilum}b we show our measurements of the dependence of
clustering on B band luminosity along with the relation found in the
2dF survey \citep{2df_clus}.  The 2dF paper interprets the luminosity
dependence as a result of the natural bias mechanism, in which
dark matter halo clustering increases with halo mass.
However, galaxies of all masses are subject to considerable clustering
dependent evolutionary forces.  Here we note that our clustering
results at $z\sim0.4$ have a luminosity dependence statistically
identical with those of the SDSS and 2dF surveys at low redshift,
confirming our ``no co-comoving co-ordinate clustering evolution''
result.  However, we also note that $r_0$ is a function of both galaxy
mass, a relative fixed property at late times, and the current star
formation rate, which can vary substantially.  The advantage of some
range of redshift is that we can use it to deepen the study of the
relation of clustering, mass, star formation and their evolution.

The theoretical predictions for clustering are clearest for dark
halos, which are increasingly clustered with mass
\citep{mowhite,jing}. The theory works quite well to predict the 
relative clustering of our small galaxy groups relative to the
clustering of individual galaxies \citep{cnoc2_gps}. However, the
luminosity dependence of clustering requires a procedure to associate
a luminosity with dark halos, ultimately requiring a fairly detailed
parameterization of the galaxy formation history. Within semi-analytic
models \citep{kauffmann,benson_a,benson_b,somerville} galaxy
clustering is expected to have a dependence on the star formation rate
and a weak luminosity dependence, but the details (even the sign of
the effect) depend significantly on the adopted cosmology, as well as
the merger and gas cooling history.  At this stage no published model
simultaneously quantitatively predicts the CNOC2 overall correlation
evolution and its star formation dependence. Some models find that the
bluest galaxies are very strongly clustered at late times, likely as a
result of a high merger rate driving star formation, which is in
significant contrast to our observational results.  Understanding the
origin of the strong color-$r_0$ relation is fundamental to
understanding the late time evolution of galaxies. Since galaxies in
small groups and pairs generate much of the correlation function,
this paper empirically investigates this question through the evolution of
galaxies in small groups.

The clustering-color (or density-star formation) relation shown in
Figure~\ref{fig:xiBR} must evolve with redshift. At the current epoch
a similar relation holds, but the bluest galaxies are a relatively
smaller fraction of the galaxy population as a consequence of ongoing
evolution.  Clustering is predicted to change slowly with
redshift. Clearly, for some fixed total stellar mass, the galaxies
with the lowest star formation rates at intermediate redshift must
have had higher than average star formation rates at high redshift. If
the clustering properties of those galaxies change slowly relative to
the rate at which their colors change, then it follows that at some
higher redshift the most strongly clustered galaxies are also the ones
with the highest star formation rates.  The Mo-White formalism
predicts that the high redshift progenitors of the galaxies with
$r_0\simeq 9 \hmpc$ must have correlations at redshifts of two or
three that increase over their lower redshift values. Even Lyman break
galaxies at $z\sim 3$ are only clustered with $r_0\simeq 4\hmpc$
\citep{adelberger}.

\section{Global Properties of Group and Field Galaxies}

The CNOC2 redshifts have a velocity precision of 70 to 100 \kms\ in
the rest frame of the galaxies. Our group selection procedure
\citep{cnoc2_gps} is applied to the $M_R^{ke}
\le -18.5$ mag sample used in the clustering analysis above.  
The CNOC2 approach to finding galaxy groups is, in brief, as
follows. Groups with three or more members are identified using an
algorithm that begins with a friends-of-friends linking in redshift
space. These trial groups are refined to a ``virialized'' set using an
iterative procedure designed to ensure that all galaxies are within
the estimated radius of virialization for a low density cosmology,
$1.5r_{200}$, where $r_{200}$ is the radius at which the mean interior
over-density is 200 times the critical value.  Kinematic group
definitions are the most objective approach available, but remain
subject to the redshift space interlopers. The resulting groups have
centers in position and redshift that do not depend strongly on the
group identification details. However, the velocity dispersion of the
groups is sensitive to the search distance in the redshift direction
and as individuals the groups are subject to large statistical errors,
see Carlberg et al. (2001) for a discussion. In spite of these
complications the statistical properties of the groups appear
to be fairly robust.

The groups used in this paper are identified with a maximum projected
distance linking length of 0.25\hmpc\ and a ``standard'' redshift
distance maximum linking length of 5\hmpc. These parameters, applied
within our group identification procedure, yield 195 groups containing
754 galaxies in the 0.1 to 0.55 redshift range. The sample is complete
to about redshift 0.45.

On the average, the groups have 3.8 members with redshifts. Given the
small numbers, there are large errors in the derived size and velocity
dispersion.  Figure~\ref{fig:Ng} shows the number of group members
with redshifts and the completeness corrected numbers versus the
estimated velocity dispersion of the groups. The total number of
groups with a given number of redshifts is given at the top of the
figure.  The numbers of group members could be increased at the cost
of more complete spectroscopy at our current $R=21.5$ mag limit, or,
by going deeper. However, the small numbers of members are fundamental
to these small groups. Similarly small groups have an unavoidable and
significant contamination from redshift space interlopers.  We have
demonstrated \citep{cnoc2_gps} that as a population these groups have
the statistical properties one would expect to find and they cannot
possibly be simply chance groupings of randomly distributed field
galaxies.

\subsection{Color-Luminosity and Color-Color Relations}

Differential evolution could lead to either completely different
galaxy populations with types non-existent in the other group, or, a
milder version in which the distribution of galaxies over type is
different, but all types are represented in both populations.  To test
for this the k-corrected colors, $(B-R)_0$, of the galaxies are
plotted against the k-corrected and evolution compensated absolute R
band absolute magnitudes for field and group galaxies in
Figure~\ref{fig:cm}.  Groups have relatively more red galaxies, but,
irrespective of environment, galaxies in both environments occupy the
same total range of colors and luminosities.  At low redshift this
effect is seen as a morphological type difference, albeit in somewhat
higher velocity dispersion groups
\citep{zab_gps}. We classify galaxies into blue and red sub-samples
using a color of $(B-R)_0=1.25$ mag. Since this color is near the
minimum of the number-color relation, it can be shifted 0.05 mag
either direction with minimal change in the two sub-samples.

The plot of the observed $U-V$ vs $V-I$ in the redshift range 0.25 to
0.45 displayed in Figure~\ref{fig:cc} further demonstrates that field
and group galaxies have essentially the same range of properties
present in different proportions. The limited redshift range allows us
to compare the colors without k-corrections.  Over-plotted are
exponential star formation models from the PEGASE.2 spectral synthesis
code at solar metallicity and no dust \citep{pegase}.  Virtually the
entire color-color plane is spanned by the exponential models. The
bluest points, which are predominantly field galaxies, clearly imply
fairly strong star formation over an extended period and require a
burst component. The reddest points can be reached from low star
formation rate models with a modest amount of dust, or, increasing the
metallicity above solar.  However, many of these points have fairly
large errors in the observed $U$ flux so may simply be scattered out
of the main distribution.

The important conclusion from these color relations is that group and
field galaxies cover the same range of properties, however the
relative frequency of blue and red galaxies is quite different in the
two environments. One interpretation is that all the galaxies have
been drawn from the same parent population, and been subject to
similar evolutionary forces, but that these have been applied to
different degrees in field and group environments. We can exclude the
possibility that the color-luminosity relations of the two populations
require unique evolutionary scenarios, even for a small subset of
their population.

\subsection{Color and Luminosity Distributions}

The color distributions, in k-corrected $(B-R)_0$, of field and group
galaxies are shown for two approximately equal-volume redshift ranges,
$z=0.15-0.36$ and $0.36-0.47$, in Figure~\ref{fig:ncol}. The sample is
restricted to galaxies with $M_R^{k}\le -20$ mag to ensure
completeness to redshift 0.55.  In this case we have not applied any
overall redshift compensation to the luminosities in order that the
redshift dependence, if present, is explicitly visible. The figure
shows that the group galaxies are on the average redder than the
field, for both redshift ranges.  The rise in the numbers (selection
function weighted) of blue field galaxies with redshift is clearly
evident. However, the blue fraction in the groups, those with
$(B-R)_0<1.25$ mag, only rises from 28\% to 31\% across the two
redshift bins, which is not significant.

Luminosity functions allow us to sharpen the comparison of group and
field galaxy evolution. We first divide the galaxies into red and blue
sub-samples. A second division of the group galaxies is required since
groups with low velocity dispersions, $\sigma_1\la 150 \kms$, are
subject to strong dynamical friction which will lead to galaxy merging
being important at a level not present in the higher velocity
dispersion groups.  We therefore split the group galaxies into two
roughly equal sub-samples at a line-of-sight group velocity dispersion
$\sigma_1=150\kms$.  Even though we do not directly volume normalize
the samples, the sample is split into the redshift ranges of 0.1--0.36
and 0.36--0.47, which yields two nearly equal volumes of about
$1.5\times 10^5 h^{-3} {\rm Mpc}^3$. The average redshifts of the two
bins are 0.26 and 0.40. The luminosity functions are calculated as the
luminosity binned sums of the magnitude selection function weights
\citep{huan_lf}, which is adequate for our purposes, although it leads
to some uncorrected incompleteness at lower luminosities at higher
redshift.

The field and two group luminosity functions are shown as summed
weights in Figure~\ref{fig:phiBR} where the thick and thin lines are,
respectively, for the low and high redshift volumes.  There is little
evolution of the red galaxy luminosity functions, for either the field
or group population. We measure a small increase in the mean
luminosity with redshift, approximately 0.1 mag, but this difference
is not statistically significant. On the other hand the blue galaxies
show about $0.3\pm0.1$ mag of increase of the mean luminosity, even
across this small redshift range. The low velocity dispersion groups
have poor statistics, but there is weak evidence that they have a
negative luminosity evolution with increasing redshift. That is, they
contain relatively more high luminosity galaxies at lower redshift.

The strong field galaxy evolution of the blue population and weak
evolution of the red population of left panels of
Figure~\ref{fig:phiBR} is a restatement of the results presented in
our more detailed paper \citep{huan_lf} where quantitative measures of
the evolution are given.  Here, our primary interest is to compare the
field galaxy evolution to the group galaxy evolution.  In the high
velocity dispersion groups evolution is largely equal to the evolution
of the field galaxies. The one possible difference is that the groups
appear to have relatively more very high luminosity red galaxies than
the field. Although this is not statistically significant in this
sample it is reminiscent of the cD galaxies found in clusters.

The important point established in this section that both the field
and $\sigma_1> 150 \kms$ group galaxy populations have very similar
evolution of their color dependent luminosity functions with
redshift. That is, in these two environments, the red galaxy
luminosity function evolves little with redshift, whereas the blue
galaxy luminosity function appears to move to higher luminosities with
increasing redshift.  Groups with velocity dispersion less than $150
\kms$ have poorer statistics and do not show evidence for any
significant luminosity function change with redshift. 

\section{Radial Color Gradients within Groups}

The radial dependence of the colors allows a test of the various
physical forces that may affect evolution. Evolutionary forces that
are universal, such as the meta-galactic ionizing flux or internal
processes of individual galaxies, should have almost no radial
dependence across a group. Environmentally dependent evolutionary
forces, such as the tidal fields of the group and its member galaxies,
or gas stripping due to an inter-group medium, should have a radial
dependence within a group. Orbital mixing will dilute the visibility
of these effects but statistically they will remain visible, if
present.

Figure~\ref{fig:colgrad} displays the radial dependence of color in
the redshift ranges $0.10-0.36$ (solid line) and $0.36-0.47$ (dashed
line), the same ranges as used for our luminosity functions. The
sample is restricted to $M_R^{ke}\le -18.5$ mag. In making these
radial plots we make no distinction between group and field galaxies,
proceeding as follows. For each group center we search the entire
volume limited sample for galaxies having $\vert \Delta v\vert \le
3\sigma_1$. For the galaxies in this sheet we calculate $r_p$, the
projected separation from the group center. We display $r_p$
normalized to the nominal virialization radius, 
$r_{200}=\sqrt(3)\sigma_1/[10H(z)]$, where $\sigma_1$ is the line of
sight velocity dispersion and $H(z)$ is the Hubble constant at the
redshift of the group
\citep{cnoc2_gps}. Here we restrict the sample to groups with
$\sigma_1\ge 150\kms$ to exclude the low velocity dispersion groups
that we consider below. The two redshift ranges have different average
k-corrected colors at large radii, a straightforward reflection of the
increasing blueness with redshift of the field population.  The mean
color of the inner $r\le 1.3r_{200}$ is $\langle (B-R)_0
\rangle=1.48\pm 0.018$ (low $z$ bin) and $\langle (B-R)_0
\rangle=1.47\pm 0.023$ (high $z$ bin) whereas at larger radii the mean color is $\langle (B-R)_0 \rangle=1.43\pm 0.005$ (low $z$ bin) and $\langle (B-R)_0
\rangle=1.35\pm 0.008$ (high $z$ bin).  We conclude the mean color difference
between field and group is highly significant.  A related effect is
seen at low redshifts in the galaxy morphologies, albeit in somewhat
richer groups \citep{tran_gps}.

The color gradient found in these groups is not as dramatic as the
equivalent measures in rich clusters \citep{balogh,pca}.  These groups
are not as ``tidy'' as clusters.  At the outer radius of the groups,
$1.5r_{200}$, the mean local volume density is about $100\rho_0$. For
a group with a velocity dispersion of 200 \kms\ the outer radius is
about 0.3\hmpc.  The redshift slice used in the color analysis would
extend 600 \kms\ or about 5\hmpc\ in either direction. Hence the
over-density of the group in that redshift slice is about 6 relative
to the unperturbed background. The same is true in clusters but the
total population is so much greater (typically at least 30 members,
instead of 3) that the overall dilution is much less in clusters.  The
probability that the galaxy is outside the group but part of the
extended group ``halo'' is about 50\% \citep{cnoc2_gps}. The halo is a
correlated structure that may well contain a color gradient relative
to the field. Hence, the projected color gradient without a background
correction is expected to be relatively unaffected by field galaxy
interlopers.  A second issue is that the velocity dispersions of the
groups, hence the calculated $r_{200}$ values have quite large
statistical errors. This will lead to some radial scrambling of the
summed group which will blur the true color gradient.  As an aside
we note that the existence of this color gradient is strong support
for the physical reality and approximately correct centers (in
$x,y,z$) of the majority of these kinematically selected groups.

A striking feature of Figure~\ref{fig:colgrad} is that the mean galaxy
colors appear to begin to become redder near $r_{200}$, the radius of
virialization. The current data are insufficient to precisely locate
the radius at which the change begins. Within the group the mean
colors do not vary much with redshift.  In these $\sigma_1>150\kms$
groups star formation is, on the average, suppressed relative to the
field.  Figure~\ref{fig:colgrad} also shows that the suppression
process operates on a timescale short compared to the overall
evolutionary timescale, since the group galaxies ``quickly'' (as
measured by group radius) reach the same colors for the two redshift
ranges examined.  The radius of virialization is where one expects
in-falling correlated structure, both associated galaxies and loosely
distributed gas, to begin to be stripped away from individual galaxies
and to spread through the group. The groups are selected to have mean
interior densities that are approximately 200 times the critical
density. This means that tidal fields, which scale as mean density
alone, will have a similar effect at all velocity
dispersions. However, ram pressure stripping, which depends on $\rho_g
\sigma_1^2$, where $\rho_g$ is the gas density, will be much less
effective in low velocity dispersion systems.  The fact that groups
with velocity dispersions near 150 \kms\ have a color gradient that
resembles that present in clusters with $\sigma \ge 800 \kms$, at
least in sign and radial range, suggests that the color change
mechanism is not related to velocity dispersion, but to density.
Tidal fields are then an acceptable mechanism to differentiate field
and group galaxies, whereas ram pressure stripping is not.

The low velocity dispersion groups have an interesting variant on the
color gradient. Color gradients of groups with maximum $\sigma_1< 100,
125$ and $150 \kms$ in the redshift range 0.25 to 0.45 are shown in
Figure~\ref{fig:col_gradhi}.  Near the centers of low velocity
dispersion groups the mean color of the high luminosity galaxy sample
becomes bluer than the field, as shown in Figure~\ref{fig:col_gradhi}.
Note that the color scale has twice the range of
Fig.~\ref{fig:colgrad}.  The remarkable result is that the low
velocity dispersion groups seem to have centers that are bluer than
the surrounding field.  The errors are based on the standard deviation
on the points in the bins. It should be noted that the inner-most bin
for the 27 low velocity dispersion groups bins contain only 3
galaxies.  These measurements suggest that in very low velocity
dispersion groups that galaxies can interact (and very likely merge)
with an accompanying burst of star formation. The sample is so small
that this result has to be viewed with caution.

\section{Galaxy Orbits in Groups}

Merging is predicted to occur in low velocity dispersion groups on a
timescale of a few group crossing times.  Theoretical studies have used
n-body simulations to examine quite a large range of idealized groups
\citep{barnes,mamon,barnes_emg,bcl,bbcl}.  Empirical evidence, such as
we have reported here (and reviewed in a low redshift context in
Hickson 1997) confirms the presence of merging in such groups.  Here
we quantitatively assess, using the parameters of our observed groups,
the impact of merging on the overall evolution of the field population
in our sample. Our groups are uniformly selected on a kinematic basis
from a field galaxy survey, so we know how to relate the group merging
to its overall impact on the galaxy population as a whole.  We use the
measured components of the positions and velocities of the observed
galaxies, and then fill in the missing components by randomly sampling
three dimensional distributions in position and velocity which we know
from our statistical study of groups
\citep{cnoc2_gps}.

Rather than full n-body integrations we follow the orbits of galaxies
in a static potential but include the dynamical friction.  We
undertake the group orbit integrations in a co-ordinate system where
$r_p$ is the co-ordinate in the plane of the sky relative to the group
center and $r_z$ is the (physical) co-ordinate along the line of
sight.  For each group galaxy we have an observed $r_p$ and $v_z$. To
complete the initial conditions we need an $r_z$ and the two
components of velocity that are in the plane of the sky.  In Carlberg
et al (2001) we found that group galaxies are distributed with a
volume density $\nu(r)$ that is consistent with the power law
$r^{-2.5}$ and that the velocity distribution is consistent with being
drawn from an isotropic Gaussian distribution.  Therefore, the two
unknown velocity components are set by drawing from a Gaussian
velocity distribution having the measured $\sigma_1$. The three
dimensional distance to the center, $r$, must lie between $r_p$ and
the largest value that keeps it within the virialized region of the
group which we take as $1.5r_{200}$, as appropriate to a low density
universe.  The density distribution $\nu(r)$ is generated from
randomly sampled numbers in $C(r|r_p)=\int_r^{r_p} \nu(x) \,dx$.  We
therefore find $r$ from $C(r| r_p) = \twothirds(r_p^{-3/2} -
r^{-3/2})$.  The line of sight position is calculated from
$r_z=\sqrt{r^2-r_p^2}$, with a randomly assigned sign. Although the
galaxy is initially within the virialized region it is possible that
the combination of radius and velocity is such that orbits move
outward for the entire duration of the integration.

The gravitational potential of the group is approximated as a static
isothermal sphere, $-2\sigma_1^2/r$, consistent with the dynamical
analysis
\citep{cnoc2_gps}. This captures the bulk of the overall dynamics
of groups, but neglects the potential fluctuations of other galaxies
orbiting in the group.  The mass of each galaxy is taken to be an
assigned $M/L$, in $\msun/\lsun$ R band, times the luminosity of the
galaxy. We use $M/L$ values of 10, 30 and 100 where $M$ is the total
mass, including dark halo, of individual galaxies. Here we present the
results only for $M/L=100$.  Integrations for $M/L=30$ have $\sim
20\%$ reduced merger rates with a factor of about 3 drop in merger
rates for $M/L= 10$.  A vital ingredient in the calculation is to
include dynamical friction.  We use Chandrasekhar's formula
\citep{bt}, adopting a Coulomb logarithm, $\ln{\Lambda}=1$. The galaxy
orbit is then integrated for a set maximum time, usually around 1Gyr.
Two example orbits over a 2 Gyr interval are shown in
Figure~\ref{fig:orbit}.

The statistics of galaxy mergers in groups are shown in
Figure~\ref{fig:mg}.  We show velocity dispersion binned merger
fractions for integrations of 0.5, 1 and 5 Gyr. Here we only count as
mergers those galaxies which simultaneously have $r<20 \hkpc$ and
$\vert v\vert \le 300\kms$ which are approximately the criteria for
merging of any galaxy with a galaxy sitting at the potential bottom of
a group. The figure shows that the only groups in which mergers will
occur with any significant rate are those with $\sigma\le 150
\kms$. The expected merger rate declines with time since the reservoir
of groups is exhausted in our simulations which have no ongoing
formation of new groups. Most of the groups that will have a merger
will do so in the first 0.5 Gyr or so.  The rate of merging over a
longer time is dependent on the creation of new groups, which is not
rapid over this interval where clustering is barely increasing
\citep{cnoc2_xi}.  The merger rate in groups, averaged over the entire
sample, is about 2.5\% per Gyr, with about 25\% of all groups below
150 \kms\ having at least one luminous galaxy merger in 1 Gyr.
Although there is no strong astrophysical justification for alternate
values of the critical velocity for merging, raising it to $400 \kms$
increases the merger rate by about 50\% and lowering it to $200 \kms$
reduces the merger rate by about a factor of four. We conclude that
groups below 150 \kms\ are short lived, although their impact on the
population as a whole to redshift one is a 15\% or so effect. The
higher velocity dispersion groups are not affected by merging at all,
so their differences from the field are due to other effects.

Figure~\ref{fig:lnsigd} shows the distribution of number of groups
with a given line-of-sight global velocity dispersions, $\sigma_1$,
for our standard groups and an alternate group sample defined using a
maximum redshift linking length of 3\hmpc. The numbers of high
velocity dispersion groups are in good agreement with the rich cluster
normalized Press-Schechter (1974) prediction. Below 200 \kms\ the
numbers of groups falls well below the prediction, mainly because they
are halos that contain only one or two galaxies above our luminosity
limit, rather than the three we require.

\section{Discussion and Conclusions}

The theme of this paper is to use the differential evolution of field
and group galaxies to examine the dominant processes that drive galaxy
evolution at low redshift. We find three empirical relations between clustering
environment and the star formation rate, as indicated by $(B-R)_0$
colors. First, the correlation length - star formation rate relation
of Figures~\ref{fig:SFR_BR} and
\ref{fig:xiBR} shows that galaxies of low star formation
rates usually occur in very strongly clustered regions.  Second, the
luminosity function evolution of the field population is very similar
to that of galaxies in groups having $\sigma_1\ga 150 \kms$,
Figure~\ref{fig:phiBR}. Third, galaxy groups with $\sigma_1\ga 150
\kms$ become redder toward their centers, Figure~\ref{fig:colgrad}.
There is a statistically weak indication that groups with low velocity
dispersions appear to have much bluer than average galaxies at the
group center.

It is natural to ask if the changes in group galaxies and the field
population as a whole are physically related.  Three physical
possibilities are outlined below.
\begin{itemize}
\item Field and group galaxies may have a common internal ``clock'' that
	causes them to evolve at similar rates. For instance, the rate
	of cooling of gas and subsequent star formation in the part of
	the galactic potential well that is occupied by stars may have
	little environmental dependence.  In this approach the
	population differences between group and field are set in place
	at some much earlier epoch.  This idea doesn't
	account for the color gradients in groups and is therefore
	unattractive.
\item Groups undergo continuing infall of field galaxies, so the evolution
	of the field is naturally mirrored in groups. In this picture
	field galaxies are where the bulk of ``normal, quiescent''
	star formation occurs. The origin of the strong evolution of
	this star formation is not specified, but, once galaxies enter
	groups their star formation rates tend to decline, as they do
	in the rich clusters. The combination of infall and subsequent
	transformation then explains both the common evolution and the
	population difference between field and group galaxy
	populations. This idea can account for color gradients, 
	but provides no insight into the mechanism driving the
	rapid evolution of blue galaxies in the field.
\item In hierarchical cosmologies all galaxies are created from smaller
	pre-existing galaxies. Moreover, many of the field galaxies in
	our sample would be found to be in groups if we counted
	galaxies further down the luminosity function. Hence, most
	galaxies can be considered to have been group galaxies at some
	time.  In this approach the natural state of weakly clustered
	galaxies is to be undergoing star formation. However, as
	galaxies enter a more clustered environment their star
	formation is reduced. The reduction is controlled by processes
	that operate on a timescale comparable to the orbital time in
	the virialized unit which is initially comparable to the
	Hubble time so the rate of reduction is slow. This picture is
	a ``two-stage'' evolutionary picture. The first stage occurs
	in regions of low velocity dispersion where there is ongoing
	star formation, accompanied by mergers from time to time that
	can lead to major star bursts. As the velocity dispersion of
	the group grows the dynamical friction timescale ascends so
	that mergers are no longer important. In the second stage the
	dominant action of the group is to tidally remove small
	gas-rich companion galaxies and more generally distributed
	gaseous material at large radii that would normally be
	available to support ongoing star formation.  Given that the
	pairwise velocity dispersion at low redshift is 300-500 \kms\
	\citep{dp,marzke,2df_vel,sdss_clus} this means that relatively
	few field galaxies are in low velocity dispersion groups where
	merging is strong. At some yet-unidentified higher redshift
	the pairwise velocity is the range where low velocity
	dispersion groups are much more prevalent and merging is
	likely to be the dominant effect of those groups. Below
	redshift one clustering has grown sufficiently that the groups
	have increased their velocity dispersion sufficiently that the
	dominant environmental evolutionary effect of groups is
	suppression of star formation by tidal effects. Given that a
	large fraction of all galaxies are found in groups and that
	clustering increases to lower redshift, the group dependent
	star formation suppression will have an increasing affect on
	the field population as a whole. Mergers are relatively rare
	over our redshift range
	\citep{patton97,patton00,cfrs_mergers,mergers}, hence
	clustering related suppression, rather than merger
	enhancement, may be the dominant source of low redshift galaxy
	evolution.
\end{itemize}

There are two observational tests that should be made of these
scenarios.  First, a general test is that if clustering is a property
that changes slowly with time, then at some redshift greater than one
there should be a population of of galaxies with $r_0\simeq 7\hmpc$,
or more, that are very strongly star forming. These would be the
progenitors of the reddest galaxies at low redshift. Second, beyond
redshift one, $\Omega_M$ begins to rise toward unity, the dark matter
clustering declines, the fraction of galaxies in groups with velocity
dispersion larger than 150 \kms\ starts to become small and the
pairwise velocity likely declines (for modest clustering biases). This
then is the time when merging in groups, possibly with much more gas
rich galaxies than observed below redshift one, leads to a dramatic
buildup of the stellar mass of spheroids through merging. Although
there are indirect suggestions of this activity a direct test with
redshifts having the kinematic precision to identify groups is
required.

\acknowledgments

This research was supported by NSERC and NRC of Canada.  HL
acknowledges support provided by NASA through Hubble Fellowship grant
\#HF-01110.01-98A awarded by the Space Telescope Science Institute,
which is operated by the Association of Universities for Research in
Astronomy, Inc., for NASA under contract NAS 5-26555.  We thank the
CFHT Corporation for support, and the operators for their enthusiastic
and efficient control of the telescope. We thank the referee and
others for suggestions that improved the presentation of this
paper.

\newpage
~

\newcounter{figi}
\newcommand{\nfig}{\addtocounter{figi}{1}\thefigi}

\figcaption[ML_SFR_BR0.ps]
{The relation between rest-frame $(B-R)_0$ and the instantaneous star
formation rate, normalized to the mass, is shown as the (green) plus
signs that falls as $(B-R)_0$ increases. The mass-to-light ratio
is shown, using the same axes, for blue (crosses) and red (circles) rest-frame
luminosities. A variety of exponential models are shown for solar and
$2\times$solar (the reddest set of model results) abundance. Points
are plotted every $3\times 10^8$ years.
\label{fig:SFR_BR}}

\figcaption[xiBR.ps]{The fitted $r_0$ (filled circles) as a function of
the k-corrected color $(B-R)_0$. The fitted slopes, $\gamma$, are
shown in the inset. The correlation lengths normalized
to $\gamma=1.8$ are shown as the pincushion symbols. The color bins
are 50\% overlapped in width.
\label{fig:xiBR}}

\figcaption[xi_lum.ps]{The fitted $r_0$ (filled circles) as a function of
the k-corrected and evolution compensated absolute magnitude (R band
in \ref{fig:xilum}a and B in \ref{fig:xilum}b) in 1 mag bins around
the central color. The fitted slopes, $\gamma$, are shown in the
inset. The correlation lengths normalized to $\gamma=1.8$ are shown as
the pincushion symbols. The open circles are the SDSS
(\ref{fig:xilum}a) and 2dF (\ref{fig:xilum}b) measurements at low redshift.
\label{fig:xilum}}

\figcaption[Ng_sig.ps]{The distribution of groups as a function
of number. The plus marks indicate the number of spectroscopic
redshifts in the identified group. The circles give the completeness
corrected numbers in the group. The groups lie in the redshift range
0.15 to 0.55. \label{fig:Ng}}

\figcaption[cm.ps]{The color vs absolute magnitude for group (red asterisk) and field (blue circle) galaxies.
\label{fig:cm}}

\figcaption[ccUVI_z38.ps]{
The observed $U-V$ versus $V-I$ colors of the group (red plus signs)
and field (blue circles) galaxies.  The lines are for a series of
exponentially declining star formation rate models with decay rates
ranging from 1~Gyr (reddest) to 6~Gyr (bluest) with star formation
starting at $t=0$.
\label{fig:cc}}

\figcaption[ncol.ps]{
The distribution of colors of field galaxies (solid line) and
group galaxies (dashed line) that are more luminous
than $M_R^{ke}=-20$ mag. The left panel is for redshifts
between 0.15 and 0.36 the right is for redshifts 0.36 to 0.47.
The error bars are simply $\sqrt{N}$ for the bins.
\label{fig:ncol}}

\figcaption[phiBR.ps]{
The k-corrected luminosity functions in the $z=0.1-0.36$ (thick lines)
and $z=0.36-0.47$ (thin lines) redshift ranges. The galaxies are
separated into field, high and low velocity dispersion groups.  These
redshift ranges have nearly equal volumes.  Blue galaxy,
$(B-R)_0<1.25$, luminosity functions are shown in the lower panels and
red galaxies in the upper panels. The luminosity functions are not
normalized to the volume, approximately $1.1\times10^5 h^{-3} {\rm Mpc}^3$.
The errors are simply $\sqrt{N}$ for the bin.
\label{fig:phiBR}}

\figcaption[col_grad.ps]{
The mean colors of galaxies as a function of projected distance from
the centers of the groups. The galaxies are selected to have
velocities with respect to the group center of $\vert \Delta v \vert
\le 3 \sigma_1$. The solid line is for galaxies with redshifts between
0.1 and 0.36 and the dashed for 0.36 to 0.47.
\label{fig:colgrad}}

\figcaption[col_gradhi.ps]{
The color-radius relation galaxies around groups having $\sigma_1<100,
$ $125,$ and $150 \kms$ (solid, dashed and dot-dashed lines,
respectively) for high luminosity galaxies with $M_R^k\le -21$. Lower
luminosity galaxies show little
 effect.
\label{fig:col_gradhi}}

\figcaption[orbit.ps]{The Monte Carlo orbits for two of the galaxies
in a random group. The box size is given in \hmpc\ units.
\label{fig:orbit}}

\figcaption[mg.ps]{The merger rate for the groups as a function
of the group velocity dispersion as derived from the Monte Carlo
integrations. The lines increase in width as the integration time
increases from 0.5, to 1.0 to 5.0 Gyr. The left panel gives the merged
fraction in each bin of velocity dispersion, whereas the right panel
gives the fraction of the entire sample that merges.
\label{fig:mg}}

\figcaption[lnsigd.ps]{The distribution of measured line-of-sight
velocity dispersions in 20\kms\ bins, plotted at the bin centers. The
asterisks (red line) are for the standard 5\hmpc\ groups and the open
circles (green line) are for the 3\hmpc\ groups. The dotted line shows
the Press-Schechter prediction for the distribution using the cluster
normalization.
\label{fig:lnsigd}}

\begin{figure}\figurenum{\nfig}
	\includegraphics{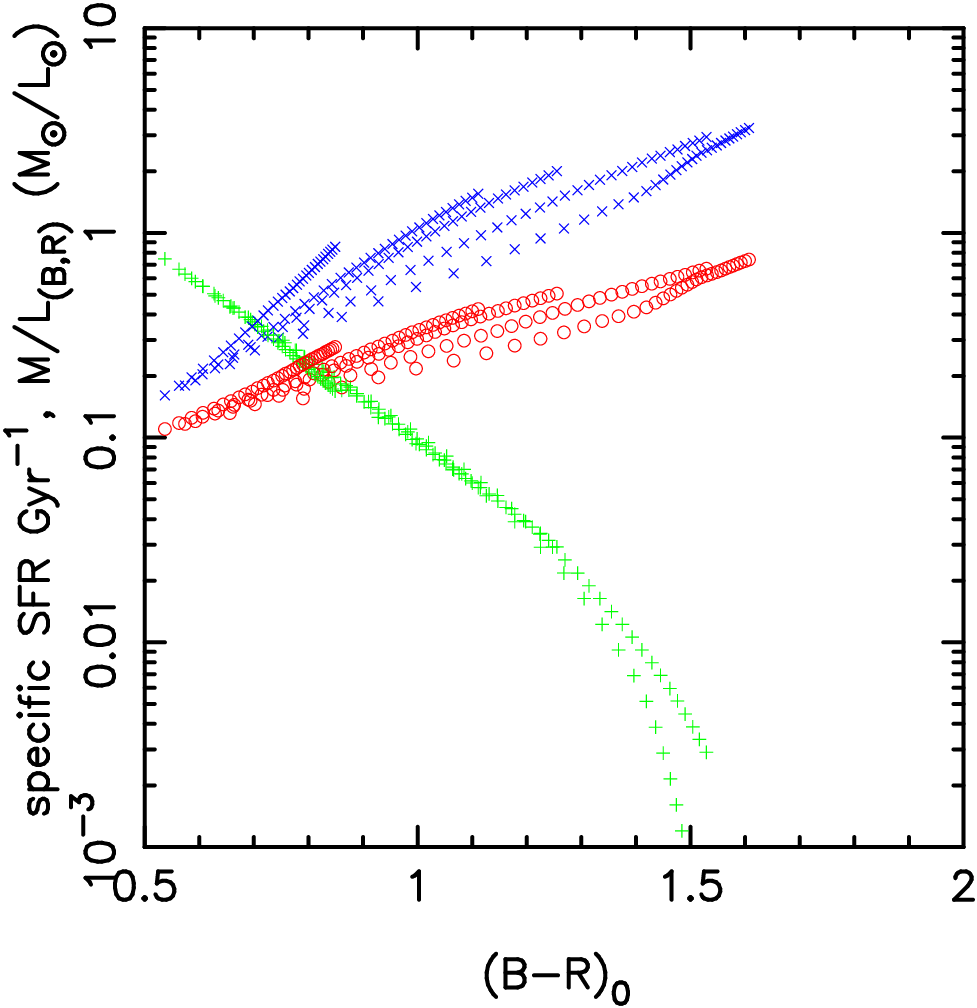} 
\caption{}
\end{figure}

\begin{figure}\figurenum{\nfig}
	\includegraphics{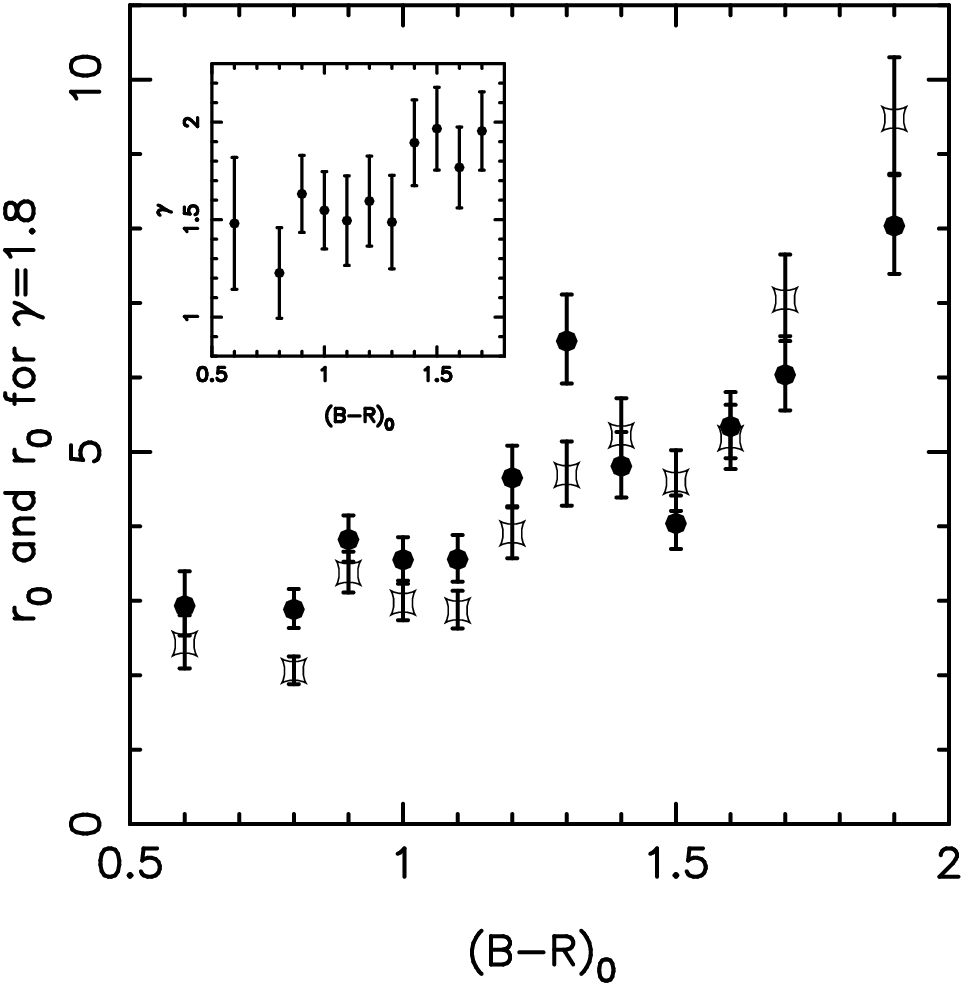} 
\caption{}
\end{figure}  

\begin{figure}\figurenum{\nfig a}
	\includegraphics{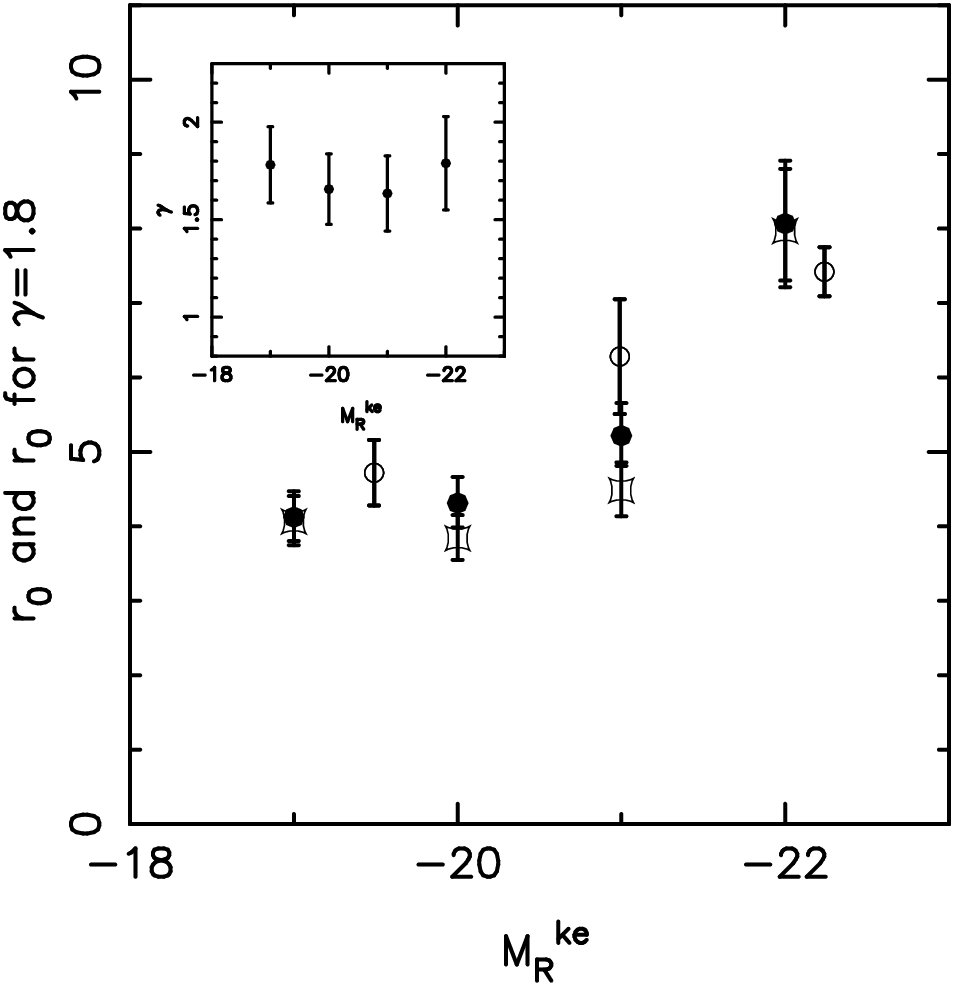} 
\caption{}
\end{figure}  
\begin{figure}\figurenum{\thefigi b}
	\includegraphics{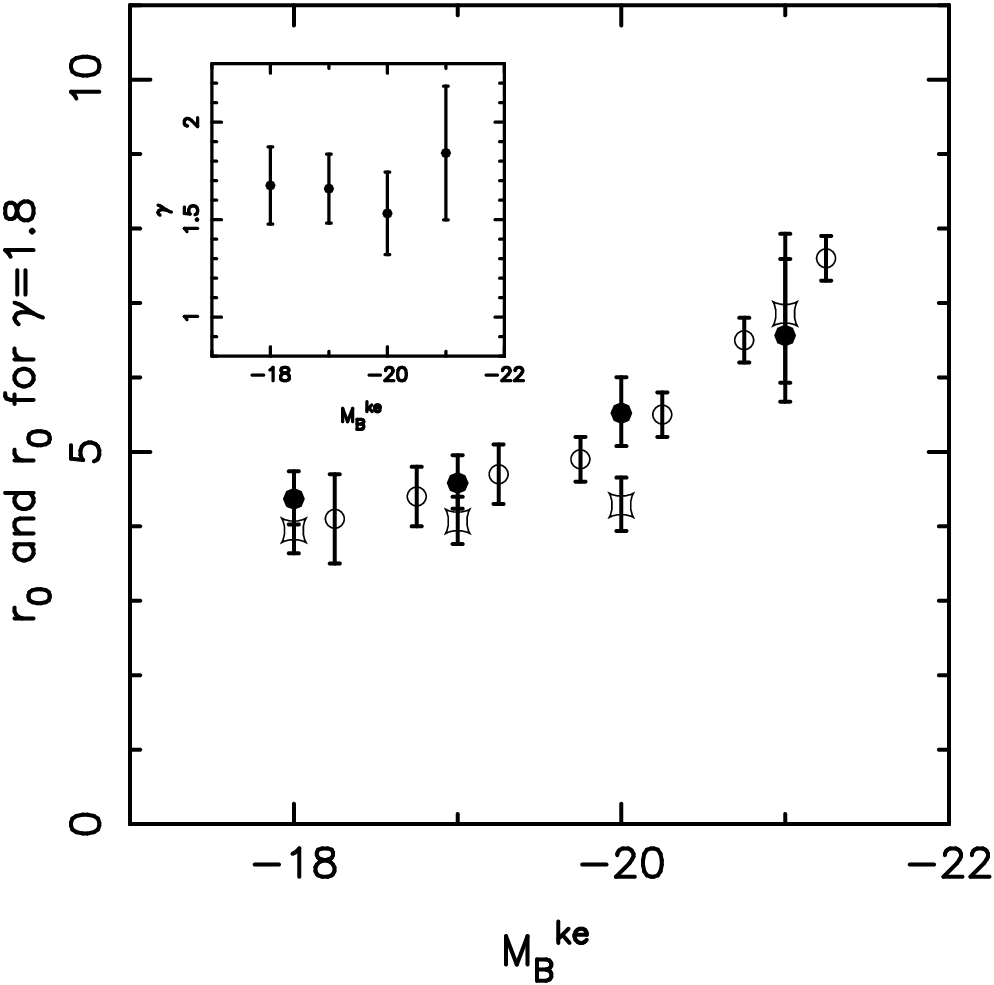} 
\caption{}
\end{figure}  

\begin{figure}\figurenum{\nfig}
	\includegraphics{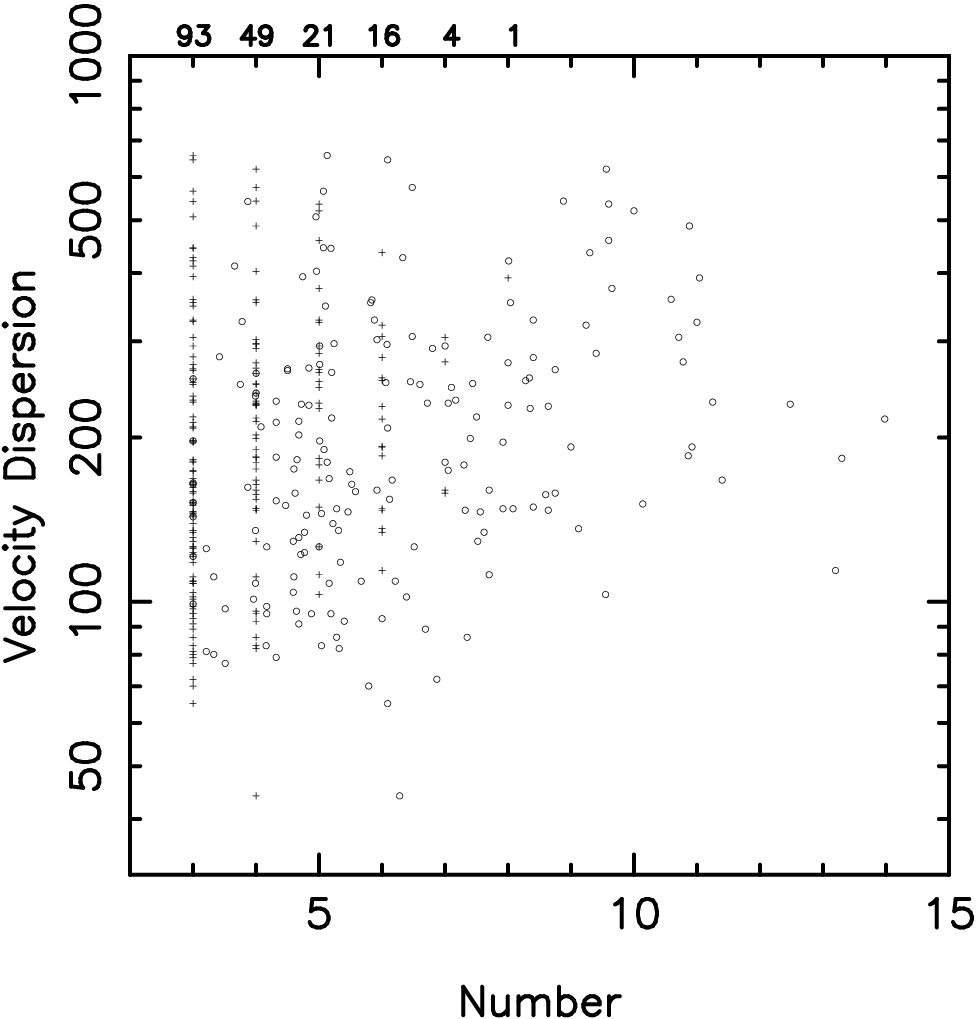} 
\caption{}
\end{figure}  

\begin{figure}\figurenum{\nfig}
	\includegraphics{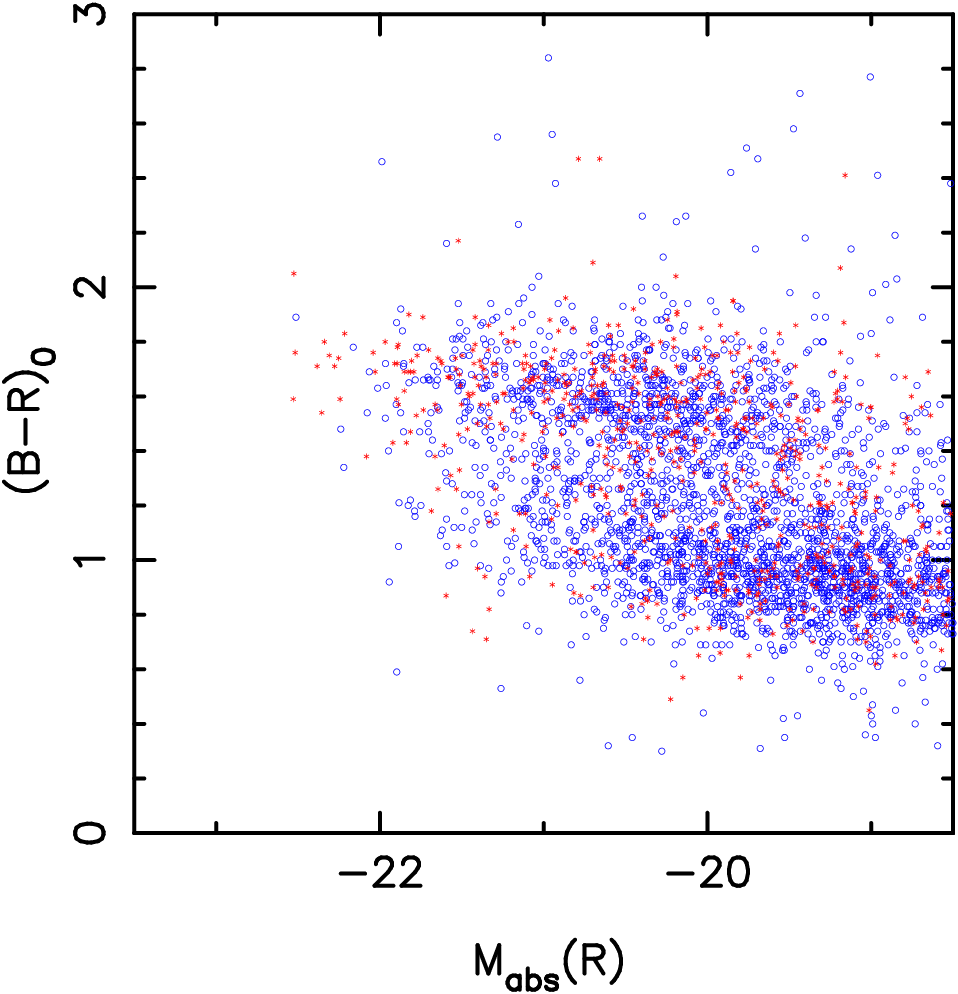} 
\caption{}
\end{figure}  
\begin{figure}\figurenum{\nfig}
	\includegraphics{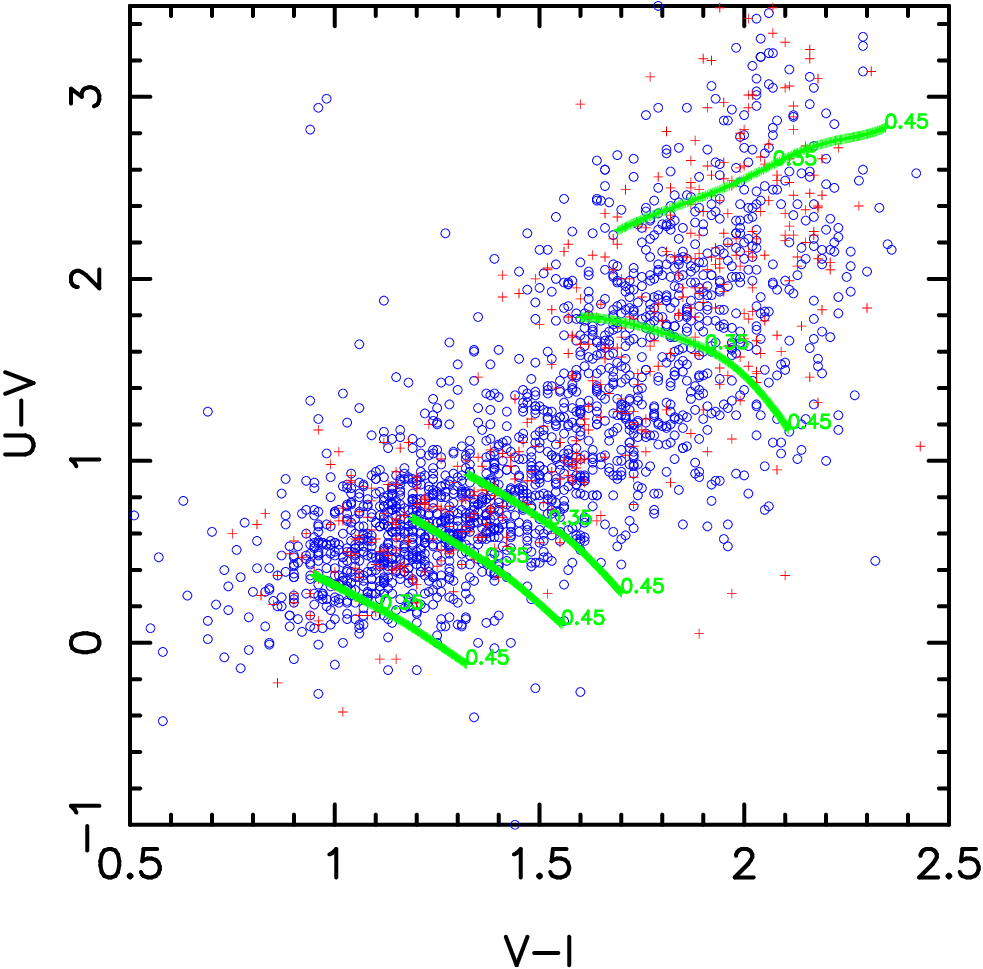} 
\caption{}
\end{figure}  

\begin{figure}\figurenum{\nfig}
\begin{picture}(0,0)(0,0)
        \put(100,-500){\includegraphics[width=0.5\hsize,angle=90]{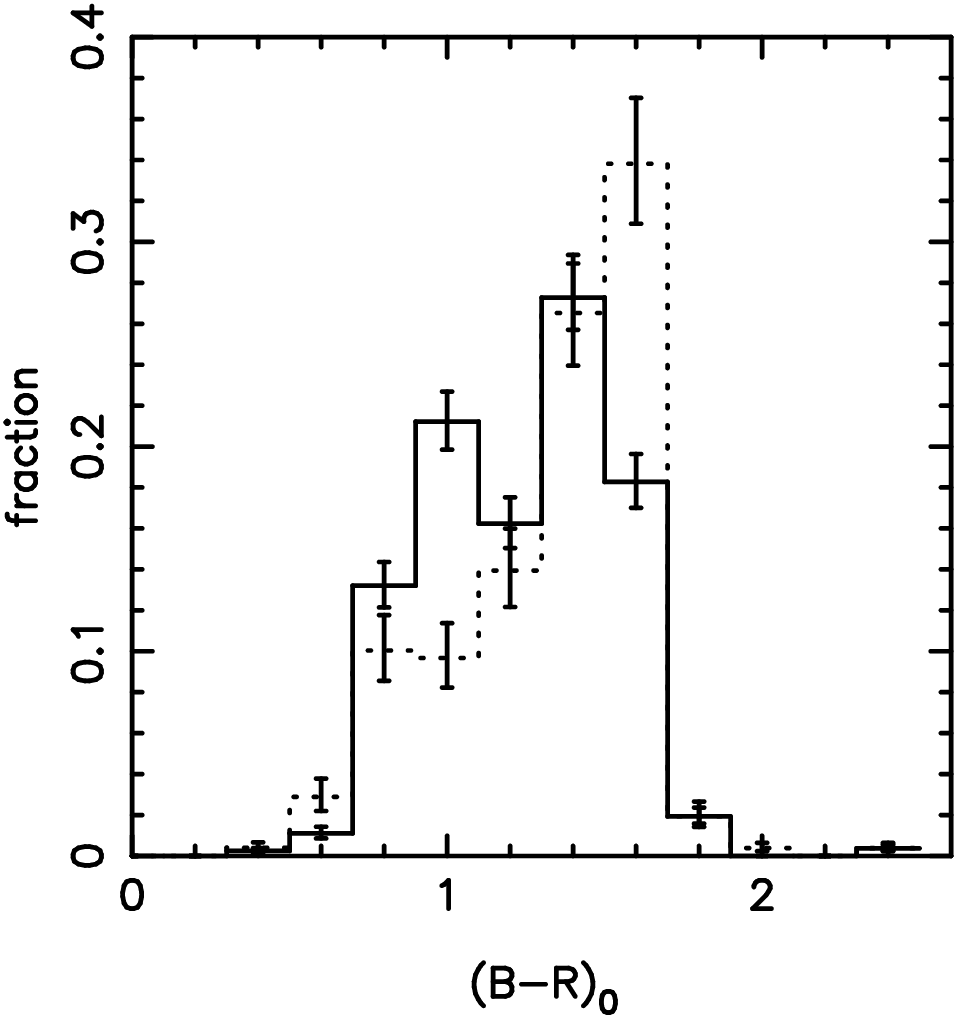}}
        \put(100,-250){\includegraphics[width=0.5\hsize,angle=90]{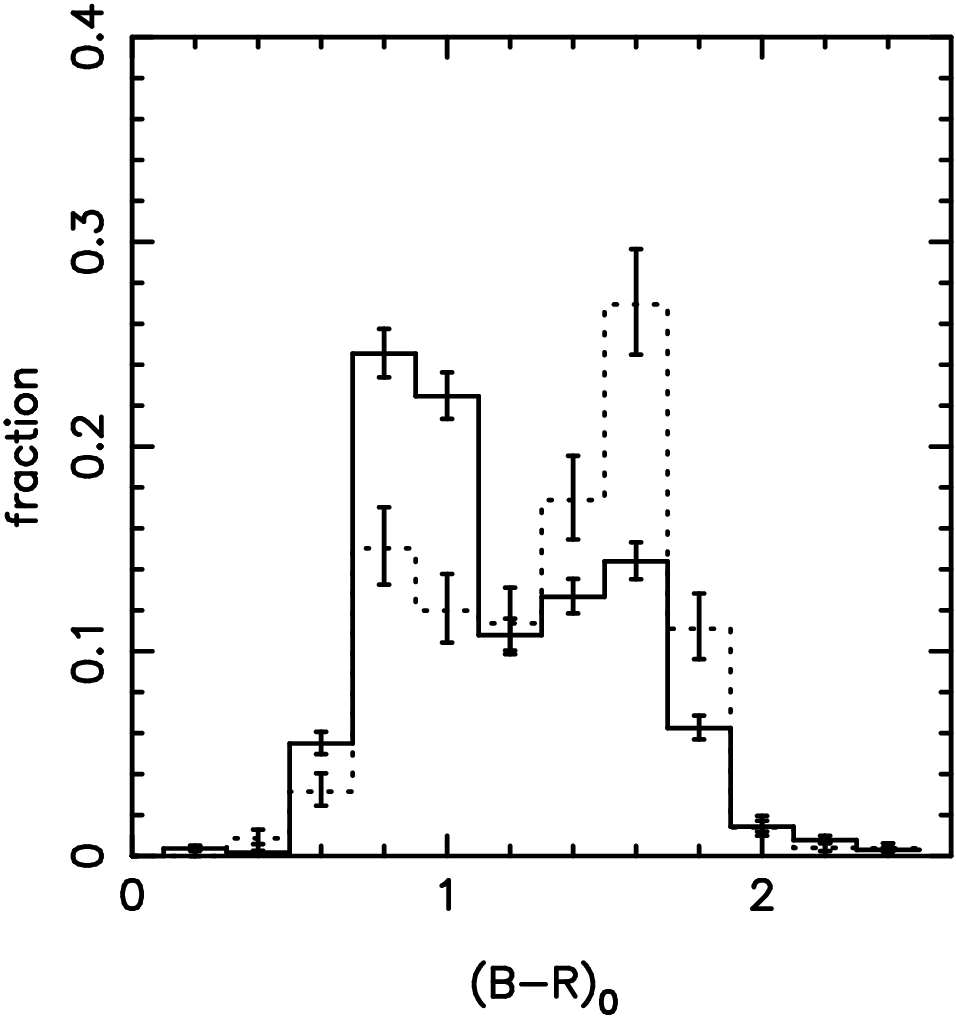}}
\end{picture}
\vspace{20cm}
\caption{}
\end{figure}  

\begin{figure}\figurenum{\nfig}
\begin{picture}(0,0)(0,0)
\put(-40,-460){\includegraphics[bb=45 28 508 498,width=0.4\hsize,angle=0]{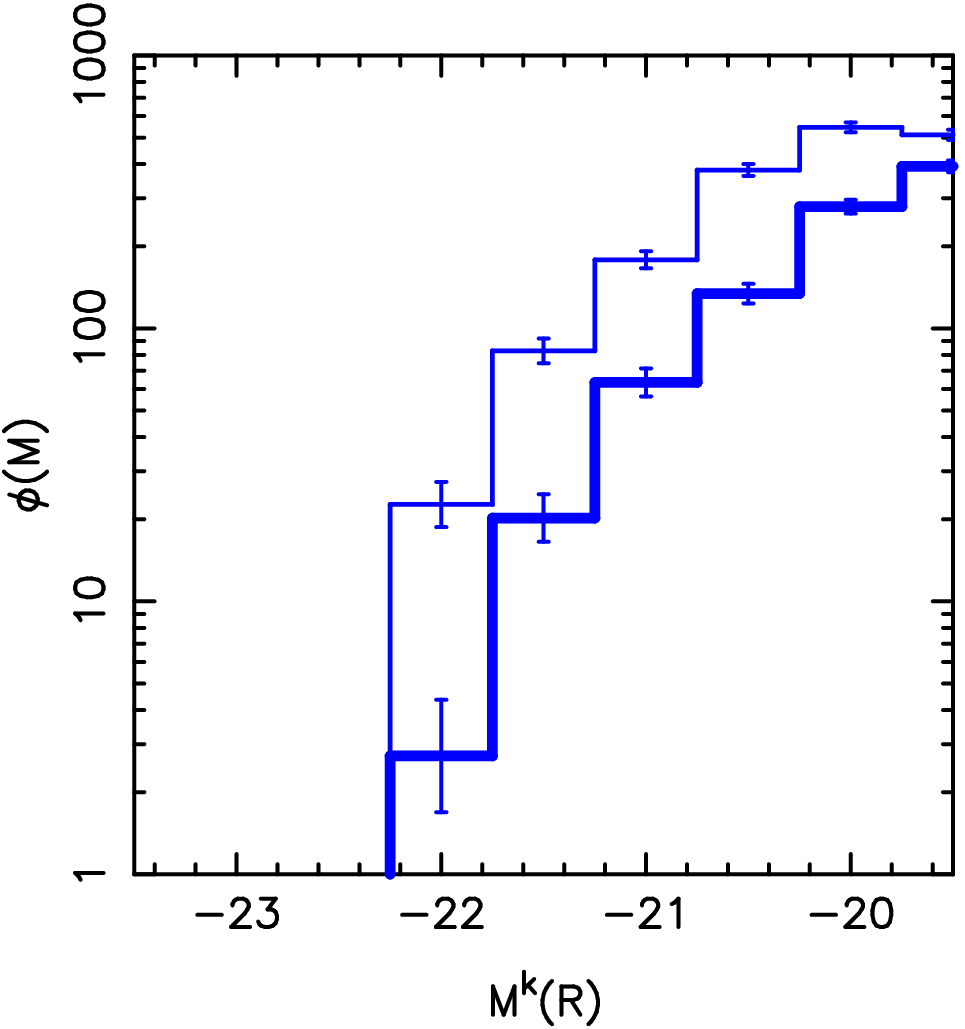}}
\put(120,-460){\includegraphics[bb=45 28 508 498,width=0.4\hsize,angle=0]{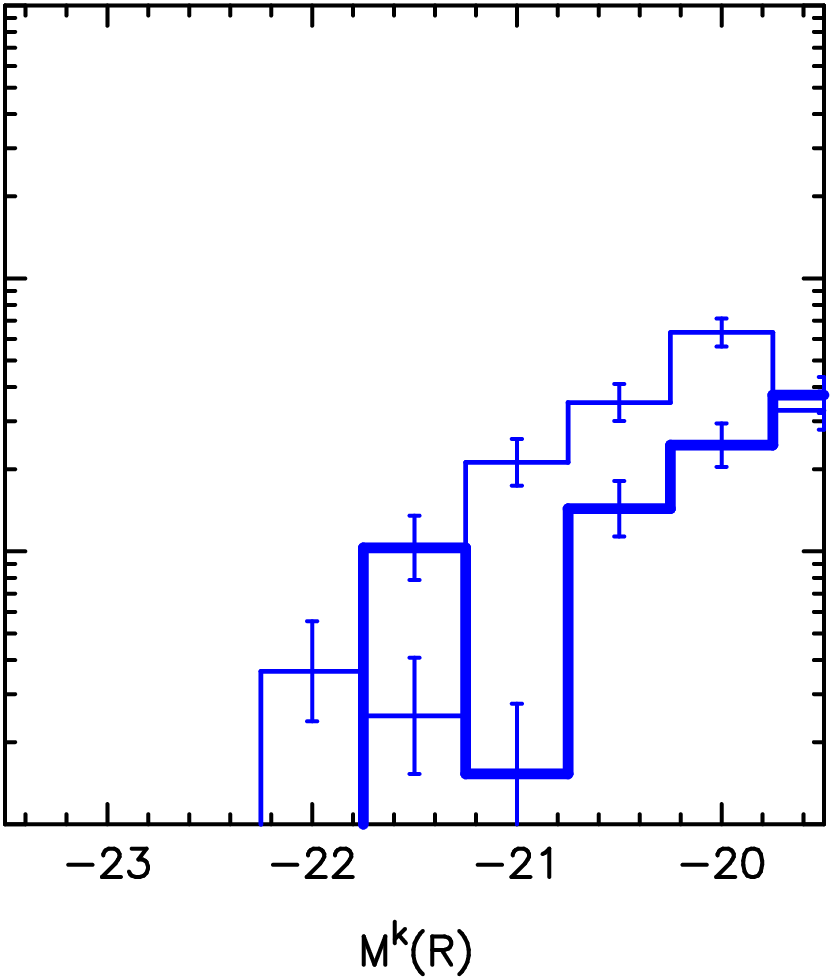}}
\put(280,-460){\includegraphics[bb=45 28 508 498,width=0.4\hsize,angle=0]{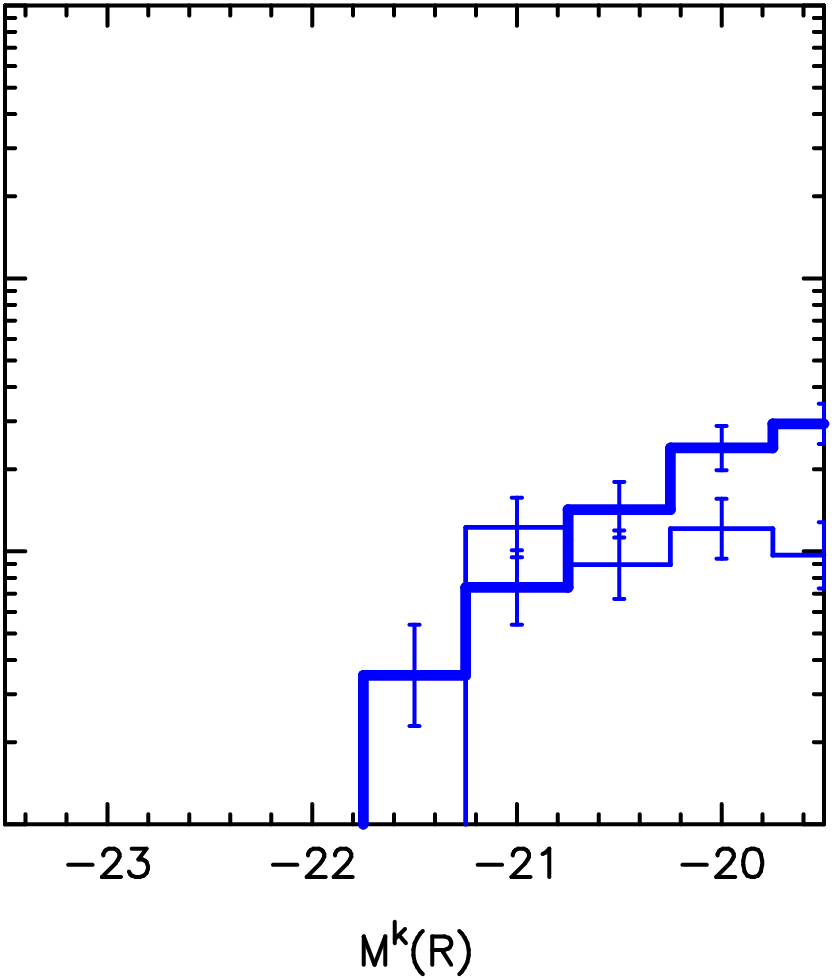}}
\put(-40,-300){\includegraphics[bb=45 28 508 498,width=0.4\hsize,angle=0]{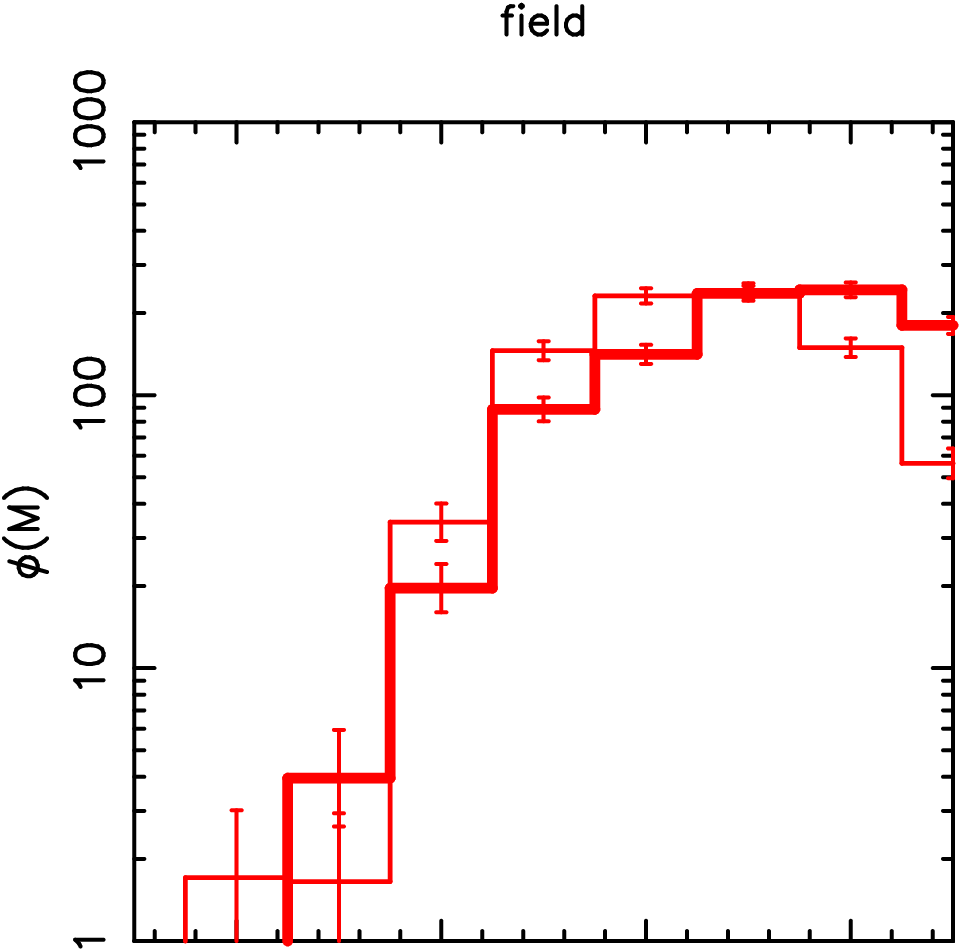}}
\put(120,-300){\includegraphics[bb=45 28 508 498,width=0.4\hsize,angle=0]{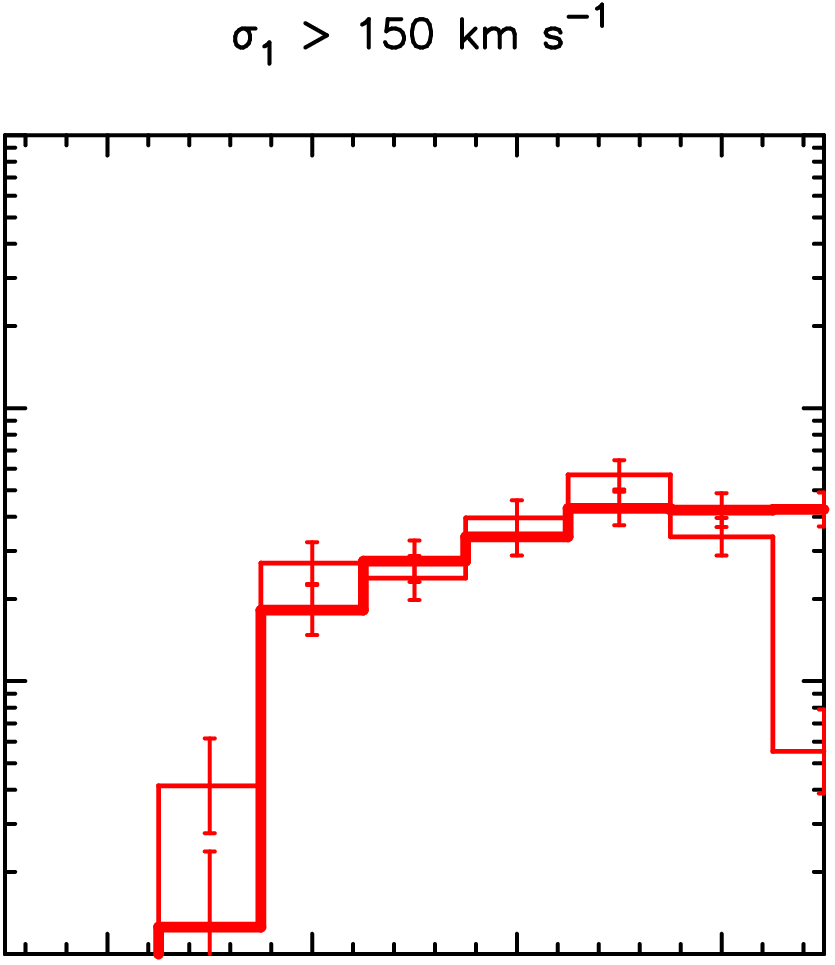}}
\put(280,-300){\includegraphics[bb=45 28 508 498,width=0.4\hsize,angle=0]{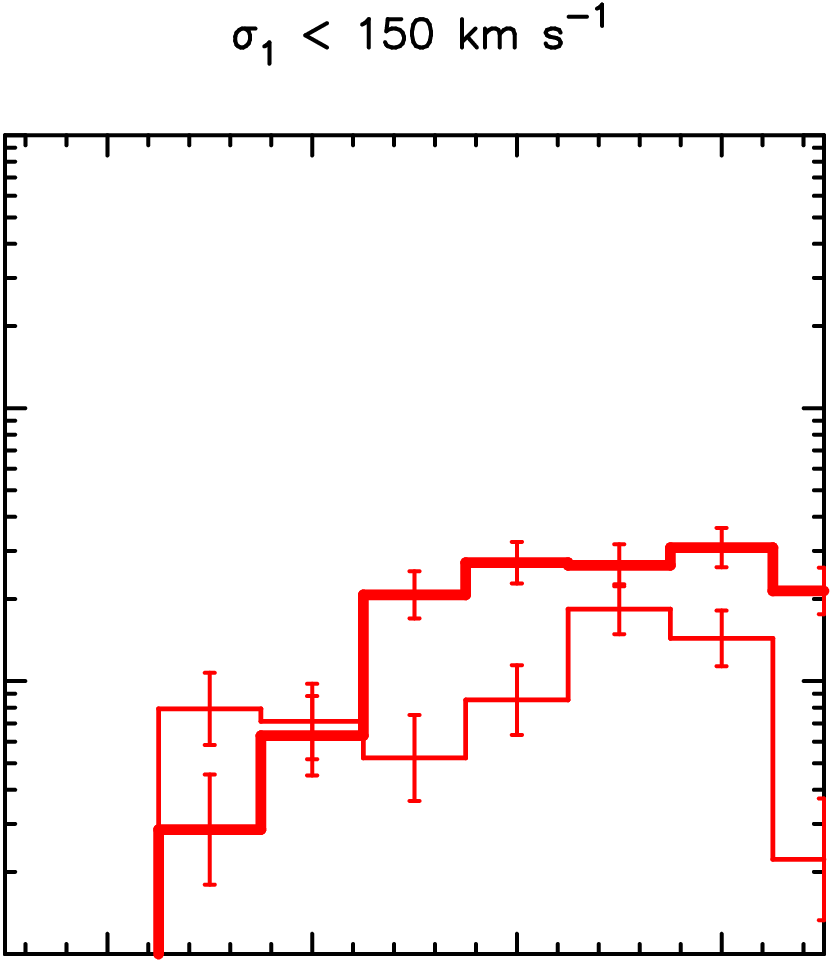}}
\end{picture}
\vspace{20cm}
\caption{}
\end{figure}  

\newpage
\begin{figure}\figurenum{\nfig}
	\includegraphics{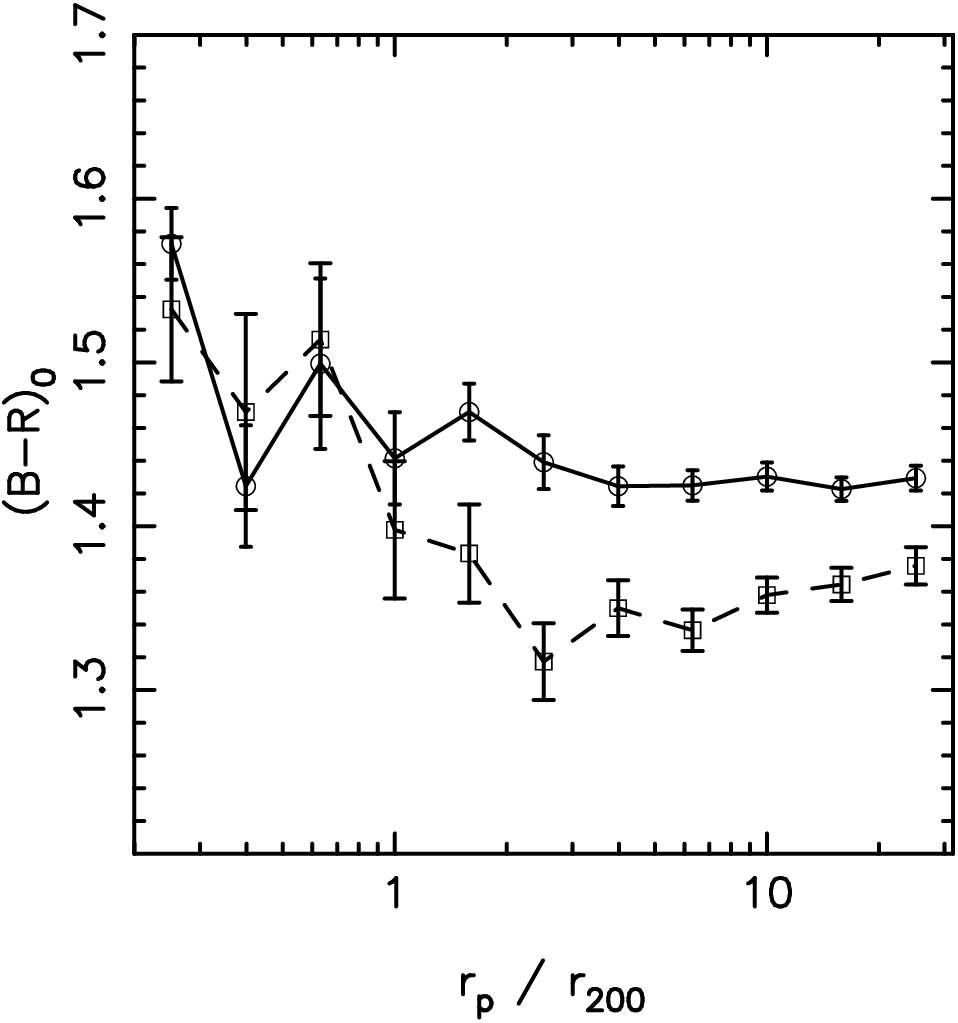} 
\caption{}
\end{figure}  

\begin{figure}\figurenum{\nfig}
	\includegraphics{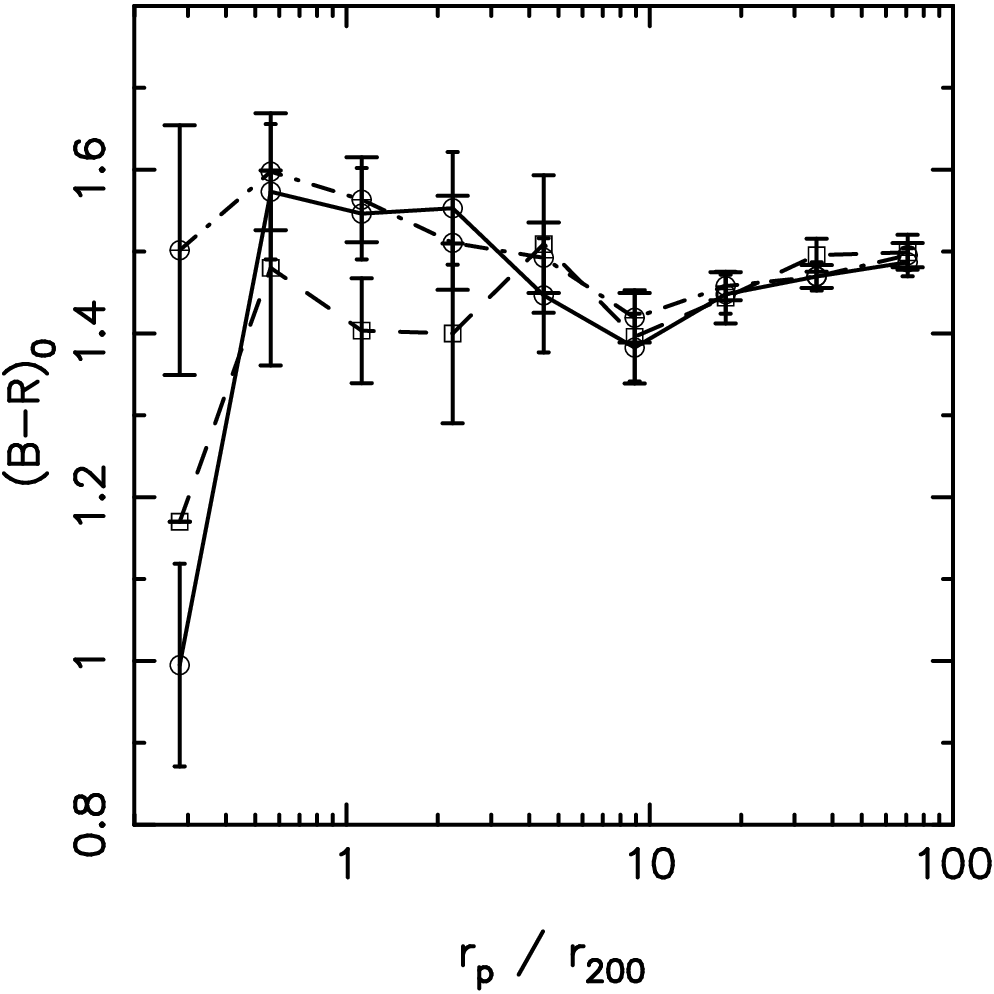} 
\caption{}
\end{figure}  

\newpage
\begin{figure}\figurenum{\nfig}
\begin{picture}(0,0)(0,0)
  \put(100,   0){\includegraphics[width=0.5\hsize,angle=-90]{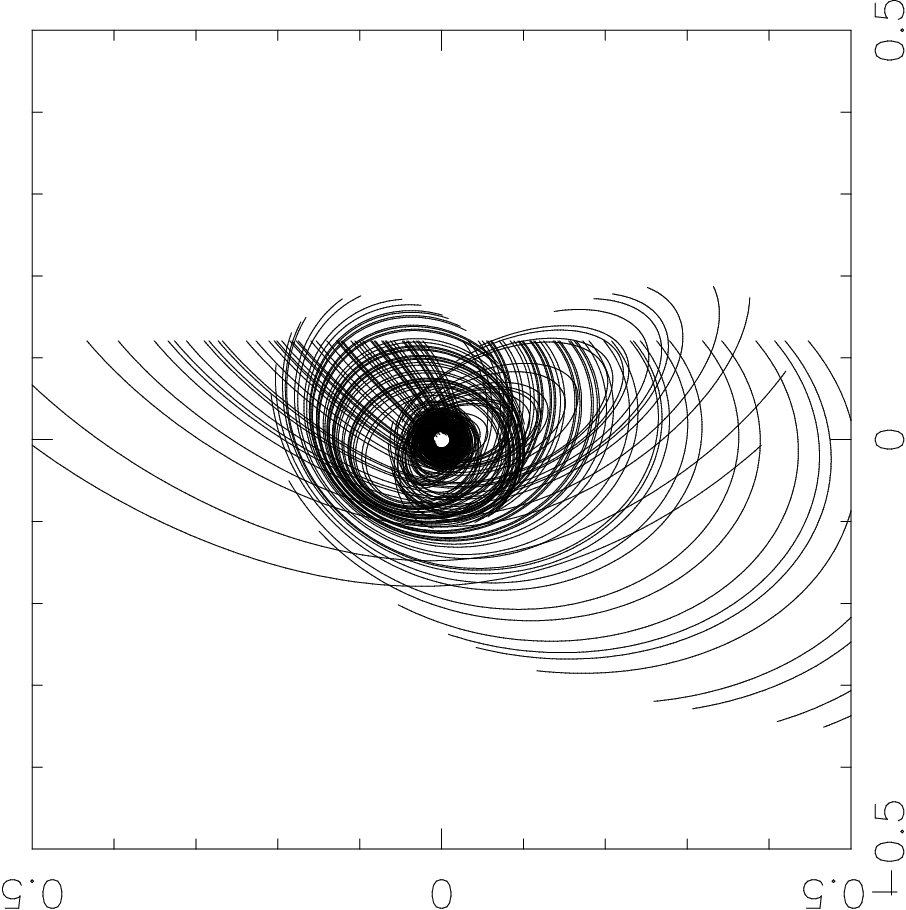}}
  \put(100,-250){\includegraphics[width=0.5\hsize,angle=-90]{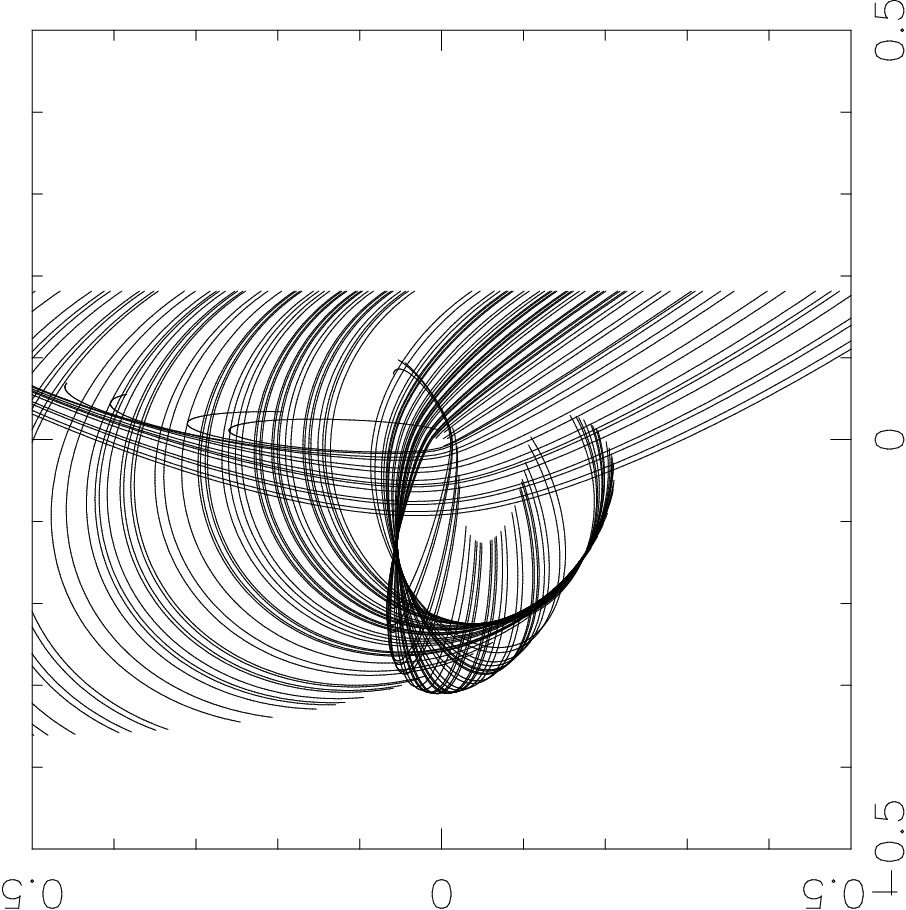}}
\end{picture}
\vspace{20cm}
\caption{}
\end{figure}  

\newpage
\begin{figure}\figurenum{\nfig}
	\includegraphics{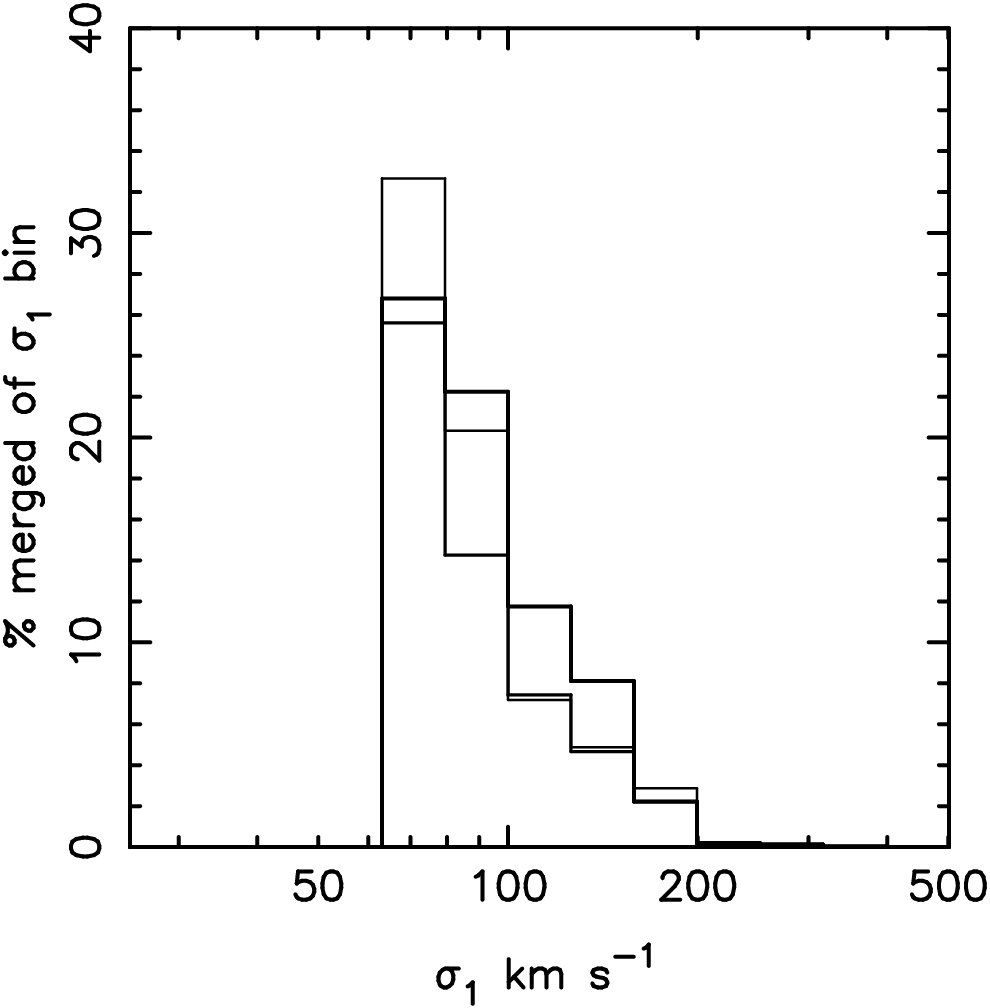}
\caption{}
\end{figure}  
\begin{figure}
	\figurenum{\nfig}
	\includegraphics{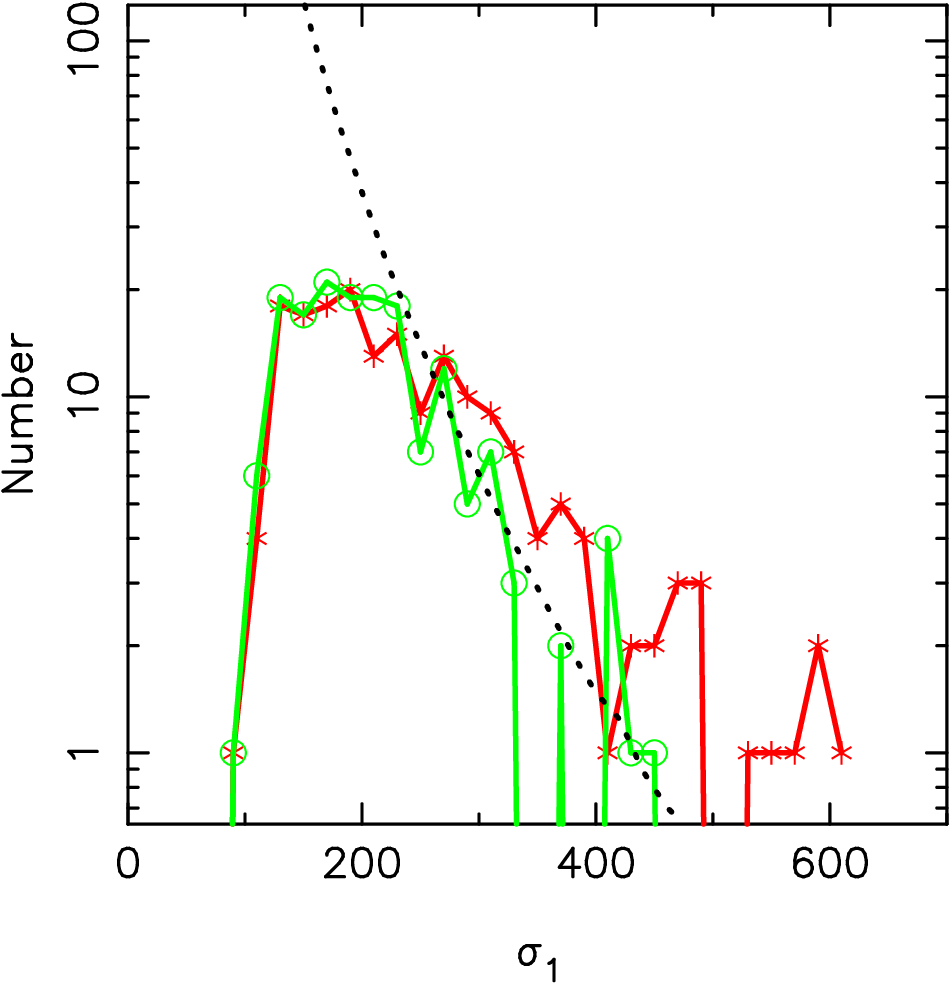} 
\caption{}
\end{figure}  

\end{document}